\documentclass[aps,preprint,superscriptaddress,nofootinbib]{revtex4}
\usepackage{amsmath}
\usepackage{graphicx}
\usepackage{amssymb}

\newcommand{\beq}{\begin{equation}}
\newcommand{\eeq}{\end{equation}}
\newcommand{\bea}{\begin{eqnarray}}
\newcommand{\eea}{\end{eqnarray}}

\newcommand{\nn}{\nonumber}

\global\long\def\fl{\mathcal{F}_{L}}
\global\long\def\fr{\mathcal{F}_{R}}

\global\long\def\gr{\mathcal{G}_{R}}
\global\long\def\ztt{Zt\bar{t}}
\global\long\def\zbb{Zb\bar{b}}
\newcommand{\gsim}{\lower.7ex\hbox{$\;\stackrel{\textstyle>}{\sim}\;$}}
\newcommand{\lsim}{\lower.7ex\hbox{$\;\stackrel{\textstyle<}{\sim}\;$}}

\begin{document}

\title{Determining $V_{tb}$ at Electron-Positron Colliders}

\author{Qing-Hong Cao}
\email{qinghongcao@pku.edu.cn}
\affiliation{Department of Physics and State Key Laboratory of Nuclear Physics
and Technology,
\\Peking University, Beijing 100871, China}
\affiliation{Collaborative Innovation Center of Quantum Matter, Beijing, China}
\affiliation{Center for High Energy Physics, Peking University, Beijing 100871, China}

\author{Bin Yan}
\email{binyan@pku.edu.cn}
\affiliation{Department of Physics and State Key Laboratory of Nuclear Physics
and Technology,
\\Peking University, Beijing 100871, China}

\begin{abstract}

Verifying $V_{tb} \simeq 1$ is critical to test the three generation assumption of the Standard Model. So far our best knowledge of $V_{tb}$ is inferred either from the $3\times 3$ unitarity of CKM matrix or from single top-quark productions upon the assumption of universal weak couplings. The unitarity could be relaxed in new physics models with extra heavy quarks and the universality of weak couplings could also be broken if the $Wtb$ coupling is modified in new physics models. In this work we propose to measure $V_{tb}$ in the process of $e^+ e^- \to t\bar{t}$ without prior knowledge of the number of fermion generations or the strength of the $Wtb$ coupling. Using an effective Lagrangian approach, we perform a model-independent analysis of the interactions among electroweak gauge bosons and the third generation quarks, i.e. the $Wtb$, $Zt\bar{t}$ and $Zb\bar{b}$ couplings. The electroweak symmetry of the Standard Model specifies a pattern of deviations of the $Z$-$t_L$-$t_L$ and $W$-$t_L$-$b_L$ couplings after one imposes the known experimental constraint on the $Z$-$b_L$-$b_L$ coupling. We demonstrate that, making use of the predicted pattern and the accurate measurements of top-quark mass and width from the energy threshold scan experiments,  one can determine $V_{tb}$ from the cross section and the forward-backward asymmetry of top-quark pair production at an {\it unpolarized} electron-positron collider. 
 
\end{abstract}

\maketitle

\section{Introduction}

The Cabibbo-Kobayashi-Maskawa (CKM) matrix element $V_{tb}$ is an important parameter in the standard model (SM) of particle physics and remains untested directly. Measuring the $V_{tb}$ accurately is important to test the unitarity of the CKM matrix and the assumption of three generations of fermions. The value of $V_{tb}$ could be modified in many 
new physics (NP) models involving extra heavy quarks~\cite{Bose:1979vz,Gronau:1984dn,Botella:1985gb,Alwall:2006bx,Kribs:2007nz,Kuflik:2012ai,Eberhardt:2012gv,Eberhardt:2012ck}. It is difficult to directly measure $V_{tb}$ experimentally. Our knowledge of $V_{tb}$ is obtained at hadron colliders either from the branching ratio of the top-quark decay into $Wb$ mode in the top-quark pair production or through measuring the single top-quark production cross sections. However, some specific assumptions have to be made in order to extract out $V_{tb}$ in both measurements.

In the first method, the top-quark decay branching ratio of the $Wb$ mode is
\beq
R = \frac{{\rm Br}(t\to Wb)}{\sum_{q=d,s,b}{\rm Br}(t \to W q)} \simeq \frac{|V_{tb}|^2}{\sum_{q=d,s,b}| V_{tq}|^2}= V_{tb}^2,
\eeq
where the $SU(2)$ coupling $g$ cancel out in both numerator and denominator. In the last step one has to use the unitarity condition of the CKM matrix, $\sum_{q=d,s,b} \left|V_{tq}\right|^2 =1$, which implicitly assumes that three and only three generations of quarks exist in the nature.  
A very high precision of $|V_{tb}|=0.99914\pm0.00005$, is derived directly from low energy precision data under the unitarity assumption of the $3\times 3$ CKM matrix~\cite{Agashe:2014kda}. 

In the second method, the single top-quark production cross sections  are proportional to weak gauge coupling $g$ and $V_{tb}$ as $\sigma_t \sim g^2 V_{tb}^2$. The limit on $V_{tb}$ can be derived under assumption of $g = g_{ Wtb}^{SM}$. The ATLAS and CMS collaborations report $\left|V_{tb}\right|=0.97^{+0.09}_{-0.10}$ and $\left|V_{tb}\right|=0.998\pm 0.038~(\rm exp.)\pm0.016~(\rm theo.)$, respectively~\cite{ATLAS:2014dja,Khachatryan:2014iya}. The gauge coupling of the $Wtb$ interaction is different from the SM prediction in several NP models, e.g. the un-unified~\cite{Georgi:1989ic,Georgi:1989xz,Hsieh:2010zr,Cao:2012ng,Cao:2015lia} and top-flavor models~\cite{Li:1981nk,Malkawi:1996fs,He:1999vp,Hsieh:2010zr,Cao:2012ng,Cao:2015lia}. In those models the third generation fermions are involved in a new gauge interaction. In such a case, one cannot link $\sigma_t$ with $V_{tb}$ directly. A precise knowledge of $V_{tb}$ will help us to extract out the gauge coupling $g$ from a precision measurement of the $Wtb$ coupling $g_{Wtb} \sim g V_{tb}$ and vice versa. It thus  offers an opportunity to verify the universality of the weak gauge coupling of SM. Observing a deviation in the gauge couplings from the SM prediction would shed light on various NP models.

Measuring $V_{tb}$ without prior knowledge of the number of fermion generations or the $SU(2)$ coupling $g$ is critical in NP searches. At an $e^+e^-$ collider with a center of mass energy $\sqrt{s}=500~{\rm GeV}$, $t\bar{t}$ pairs would be copiously produced, with several 100,000 events for an integrated luminosity of $500~{\rm fb}^{-1}$~\cite{Amjad:2013tlv}.  Such a large dataset offers a great opportunity of testing top-quark properties. 
In this work we perform a model-independent analysis of the gauge interaction of the third generation quarks.  We argue that one could determine the $V_{tb}$ from the precision measurements of the top-quark pair  production at an {\it unpolarized} $e^+ e^-$ collider with $\sqrt{s}\sim 350-1000 {\rm GeV}$. Furthermore, we show that the measurement of $V_{tb}$ is not sensitive to the collider energy in our method and $\sqrt{s}=500$ GeV is enough to measure $V_{tb}$ at percentage level.

\section{Effective gauge couplings}

So far no heavy resonances are observed at the LHC yet. It is reasonable to assume the NP effects modify the SM theory prediction slightly and can be described by a set of higher dimensional operators made out of the SM fields~\cite{Buchmuller:1985jz},
\bea
\mathcal{L}_{\rm eff}=\mathcal{L}_{\rm SM}+\frac{1}{\Lambda^2}\sum_i (c_i\mathcal{O}_i+h.c.)+\mathcal{O}\left(\frac{1}{\Lambda^3}\right),
\eea
where $\mathcal{L}_{\rm SM}$ denotes the Lagrangian of the SM, $\Lambda$ is the characteristic scale of NP, $\mathcal{O}_i$ is the dimension-6 operator which satisfies the $SU(3)_c\otimes SU(2)_L\otimes U(1)_Y$ gauge symmetry, and $c_i$ is Wilson coefficient which represents the strength of the operator $\mathcal{O}_i$. 
In this work, we consider those operators affecting top-quark gauge couplings and restrict ourselves to the interference between the SM and the operators when computing the effects of NP operators.
Since the left-handed top quark and bottom quark form a $SU(2)_L$ weak doublet, the $W$-$t_L$-$b_L$ coupling is always related to the $Z$-$t_L$-$t_L$ and $Z$-$b_L$-$b_L$ couplings~\cite{Berger:2009hi}. It is, therefore, reasonable to combine the tree-level induced effective operators which are related to those couplings to determine $V_{tb}$,~\footnote{The loop induced operators are usually suppressed by $1/(16\pi^2)$ and not considered in our analysis. }
\begin{align}
\mathcal{O}_{\phi q}^{(1)} &= i\left(\phi^{\dagger}D_{\mu}\phi\right)\left(\bar{q}\gamma^{\mu}q\right), &
\mathcal{O}_{\phi q}^{(3)} &= i\left(\phi^{\dagger}\tau^{I}D_{\mu}\phi\right)\left(\bar{q}\gamma^{\mu}\tau^{I}q\right), \nonumber \\
 \mathcal{O}_{\phi t} &= i\left(\phi^{\dagger}D_{\mu}\phi\right)\left(\bar{t}_{R}\gamma^{\mu}t_{R}\right), &
\mathcal{O}_{\phi b} &= i\left(\phi^{\dagger}D_{\mu}\phi\right)\left(\bar{b}_{R}\gamma^{\mu}b_{R}\right), \nonumber\\
\mathcal{O}_{\phi\phi} &= i\left(\tilde{\phi}^{\dagger} D_{\mu}\phi\right)\left(\bar{t}_{R}\gamma^{\mu}b_{R}\right), &&
\end{align}
where $D_{\mu}=\partial_{\mu}-ig(\tau^I/2)W_{\mu}^I-ig^{\prime}YB_{\mu}$  is the gauge-covariant derivative, $g$ and $g^\prime$ are the gauge couplings of $SU(2)_L$ and $U(1)_Y$, respectively, and $Y$ is the hypercharge of the field to which $D_{\mu}$ is applied, $\tau^I$ is the usual Pauli matrix; $q^T=(t_{L},\, b_{L})$ is the left-handed top-bottom $SU(2)_{L}$ doublet; $t_{R}(b_{R})$
are corresponding to the right-handed isosinglets; and $\phi$ is $SU(2)_L$ weak doublet of Higgs field, defined as $\phi^T=1/\sqrt{2}~(0,v+h)$ with $v=246$ GeV in the unitarity gauge with $\tilde{\phi}=i\tau^2\phi^{*}$. 
After symmetry breaking $\langle\phi\rangle=v/\sqrt{2}$,
the set of operators generates the following corrections to
the couplings $Wtb$, $\ztt$ and $\zbb$~\cite{Yang:1997iv,Whisnant:1997qu,AguilarSaavedra:2008zc,Berger:2009hi},
\begin{eqnarray}
\mathcal{O}_{Wtb} & = & \frac{c_{\phi q}^{(3)}v^{2}}{\Lambda^{2}}\frac{g}{\sqrt{2}}W_{\mu}^{+}\bar{t}_{L}\gamma^{\mu}b_{L}+\frac{c_{\phi\phi}v^{2}}{2\Lambda^{2}}\frac{g}{\sqrt{2}}W_{\mu}^{+}\bar{t}_{R}\gamma^{\mu}b_{R}+h.c.,\label{eq:wtb}\\
\mathcal{O}_{\ztt} & = & \frac{\left(c_{\phi q}^{(3)}-c_{\phi q}^{(1)}\right)v^{2}}{\Lambda^{2}}\frac{g}{2 c_W}Z_{\mu}\bar{t}_{L}\gamma^{\mu}t_{L}-\frac{c_{\phi t}v^{2}}{\Lambda^{2}}\frac{g}{2 c_W}Z_{\mu}\bar{t}_{R}\gamma^{\mu}t_{R},\label{eq:ztt}\\
\mathcal{O}_{\zbb} & = & -\frac{\left(c_{\phi q}^{(1)}+c_{\phi q}^{(3)}\right)v^{2}}{\Lambda^{2}}\frac{g}{2 c_W}Z_{\mu}\bar{b}_{L}\gamma^{\mu}b_{L}-\frac{c_{\phi b}v^{2}}{\Lambda^{2}}\frac{g}{2 c_W}Z_{\mu}\bar{b}_{R}\gamma^{\mu}b_{R},\label{eq:zbb}
\end{eqnarray}
where $c_W\equiv \cos\theta_W$ is the cosine of the weak mixing angle.

The anomalous coupling $\gr\equiv c_{\phi\phi}v^2/(2\Lambda^2)$ in the $Wtb$ coupling, as severely constrained by the $b\to s\gamma$ data~\cite{Grzadkowski:2008mf,Cao:2015doa}, is within the window of  $-8\times 10^{-4} \leq \gr \leq 2.1 \times 10^{-3}$. 
Furthermore, the LEP precision measurements require a strong cancellation between the two operators $\mathcal{O}_{\phi q}^{(1)}$ and $\mathcal{O}_{\phi q}^{(3)}$, i.e. $c_{\phi q}^{(1)}+c_{\phi q}^{(3)}\simeq 0$, which leaves the SM $Z$-$b_L$-$b_L$ coupling unmodified~\cite{Abdallah:2008ab}. 
It immediately enforces a correlation among the deviations of $W$-$t_L$-$b_L$ coupling and $Z$-$t_L$-$t_L$ coupling as follows:
\bea
 g_{ Wtb}^{\rm NP} &=& (\Delta V_{tb}+ \mathcal{F}_L)\frac{g}{\sqrt{2}}W^+_\mu\bar{t}_L\gamma^\mu b_L+h.c., \nonumber \\ 
 g_{ Ztt}^{\rm NP}  &=& 2 \mathcal{F}_L \frac{g}{2c_W}Z_\mu \bar{t}_L \gamma^\mu t_L +  \mathcal{F}_R\frac{g}{2c_W}Z_\mu \bar{t}_R \gamma^\mu t_R,
\eea
where $\Delta V_{tb}$ is the deviation of the $V_{tb}$ matrix element from the SM value $V_{tb}^0= 1$, $\mathcal{F}_L=c_{\phi q}^{(3)} v^2/\Lambda^2$ and $\mathcal{F}_R = - c_{\phi t} v^2/\Lambda^2$. We assume the three coefficients are real in our calculation. Throughout this work $\Delta X \equiv X-X^0$ represents the deviation of the central value of variable $X$ from the theory prediction $X^0$ and $\delta X$ denotes the experimental error of $X$.
Notice the relation between the coefficients of the left-handed neutral and charged currents~\cite{Malkawi:1994tg,Carlson:1994bg,Berger:2009hi},
\bea
 (g_{Ztt}^{\rm NP})_L=2\mathcal{F}_L=2(g_{Wtb}^{\rm NP})_L-2\Delta V_{tb}.
\eea
The relation holds for any underlying theory with an approximate custodial symmetry such that the vertex $Z$-$b_L$-$b_L$ is not modified as discussed above~\cite{Malkawi:1994tg}. It is possible to yield such a twice factor from an additional symmetry, e.g. certain subgroups of the custodial symmetry which protect the $\rho$ parameter~\cite{Sikivie:1980hm} and the $Z$-$b_L$-$b_L$ coupling~\cite{Agashe:2006at}.

Recently, measuring $Wtb$ anomalous couplings at the Large Hadron Collider (LHC) with recent experimental data are studied in Refs.~\cite{Chen:2005vr,Cao:2007ea,Fabbrichesi:2014wva,Bernardo:2014vha,Cao:2015doa}, which shows $-0.06\leq\Delta V_{tb}+ \fl\leq 0.03$ at 95\% confidence level. The $\ztt$ coupling can be measured from the associated production of the top-quark pair and $Z$-boson. It is shown in Ref.~\citep{Rontsch:2014cca} that $-0.99\leq\fl\leq 0.57$ at 95\% confidence level at the 13~TeV LHC with 300 ${\rm fb}^{-1}$. To further constrain the $\fl$ at the LHC, it demands a fairly large luminosity to achieve a good precision. It is impossible to obtain an accurate $V_{tb}$ from the $Wtb$ and $Zt\bar{t}$ measurement at the LHC.  The electron-positron collider provides a great opportunity to precisely determine both $\mathcal{F}_L$ and $V_{tb}$ in the top-quark pair production.

\section{Precisions at the $e^+e^-$ collider}
\label{eepre}

\subsection{Top-quark width measurement}

At the $e^+ e^-$ collider the $Wtb$ coupling can be extracted out from the top-quark width ($\Gamma_t$) measurements around the threshold region of a pair of top quarks~\cite{Horiguchi:2013wra}. 
In the SM the top-quark entirely decays into a bottom-quark and a $W$-boson. A state-of-art calculation of the top-quark decay width at next-to-next-to-leading order in quantum chromodynamics,  including next-to-leading order electroweak correction and finite bottom quark mass and $W$ boson width effect, is carried out recently in Ref.~\cite{Gao:2012ja}, which shows the top-quark width in the SM is
\beq
\Gamma_t^{0}\equiv \Gamma_t^{\rm NNLO} = 0.8926 \times \Gamma_t^{\rm LO},
\eeq  
where $\Gamma_t^{\rm LO}$ labels the top-quark decay width at the leading order in the limit of $m_{b,s,d} \to 0$,
\beq
\Gamma_t^{\rm LO} = \frac{G_F m_t^3}{8\sqrt{2}\pi} \sum_{i=d,s,b}\left| V_{ti}\right|^2 \left(1-\frac{m_W^2}{m_t^2}\right)^2\left(1+\frac{2m_W^2}{m_t^2}\right).
\eeq
Here, $m_t$ denotes the top-quark mass, $m_W$ is the $W$-boson mass and $G_F=1.166\times 10^{-5}\rm{GeV}^{-2}$~\cite{Agashe:2014kda}.
Using the branching ratio measurement, 
\beq
{\rm Br}(t\to Wb) = \frac{\Gamma(t\to Wb)}{\sum_{i=d,s,b}\Gamma(t\to W i)}=\frac{|V_{tb}|^2}{\sum_{i=d,s,b}|V_{ti}|^2},
\eeq
we obtain
\beq
\Gamma_t^{\rm LO} = \frac{G_F m_t^3}{8\sqrt{2}\pi} \frac{\left| V_{tb}\right|^2}{{\rm Br}(t\to Wb)} \left(1-\frac{m_W^2}{m_t^2}\right)^2\left(1+\frac{2m_W^2}{m_t^2}\right).
\eeq
Here, we assume ${\rm Br}(t\to Wb)$ is the same as the SM prediction. Deviations of $g$, $m_t$ and $V_{tb}$ from the SM values modify the top-quark width as following
\beq
\frac{\Delta \Gamma_t}{\Gamma_t^{0}} =  3 \frac{\Delta m_t}{m_t} + 2 \Delta V_{tb} + 2 \mathcal{F}_L,
\label{eq:dwidth}
\eeq
where $\Delta \Gamma_t\equiv\Gamma_t-\Gamma_t^{0}$.
Both $m_t$ and $\Gamma_t$ can be measured precisely from the threshold scan at the $e^+e^-$ collider with $\sqrt{s}=340-350~{\rm GeV}$. It is shown that $m_t$ and $\Gamma_t$ could be measured with an accuracy of 0.006\% and 0.5\%, respectively~\cite{Gomez-Ceballos:2013zzn}. Thus, we ignore the small deviation of the top quark mass and assume 
the error $\delta \Gamma_t/\Gamma_t^{0} \simeq 0.01$ hereafter. Under such a circumstance $\Delta V_{tb}$ depends mainly on the precision measurement of $\mathcal{F}_L$,
\bea
\Delta V_{tb}\simeq \frac{1}{2}\frac{\Delta\Gamma_t}{\Gamma_t^{0}}-\mathcal{F}_L.
\label{eq:dvtb}
\eea 
One then can determine the $V_{tb}$ if the $\mathcal{F}_L$ could be measured precisely. 
However, it is difficult to measure $\mathcal{F}_L$ from the $Z$-boson and top-quark pair associated production ($pp \to Z t\bar{t}$) at the LHC~\cite{Baur:2004uw,Baur:2005wi,Rontsch:2014cca}. On the other hand, $\mathcal{F}_L$ could be well measured at a $e^+e^-$ collider through the process of $e^+ e^- \to \gamma/Z \to t \bar{t}$~\cite{Baer:2013cma,Amjad:2013tlv,Asner:2013hla,Barducci:2015aoa,Amjad:2015mma}. In this study we focus on unpolarized electron and positron beams. 

\subsection{Top-quark pair production}

The anomalous couplings of $\mathcal{F}_L$ and $\mathcal{F}_R$ can be measured from the inclusive cross section of $t\bar{t}$ pair produciton ($\sigma_{t\bar{t}}$) and the forward-backward asymmetry of top quarks ($A_{FB}$), which is defined as 
\beq
A_{FB} \equiv \frac{\sigma_F -\sigma_B}{\sigma_F +\sigma_B},
\eeq
where 
\beq
\sigma_F = \int_0^1 \frac{d\sigma}{d\cos\theta_t} d\cos\theta_t~,~\sigma_B = \int_{-1}^0 \frac{d\sigma}{d\cos\theta_t} d\cos\theta_t.
\eeq
with $\theta_t$ being the polar angle of top-quark inside the center-of-mass frame. 
A simple algebra shows
\bea
\label{eq:cf}
\sigma_{t\bar{t}} &=& \sigma^{ 0}_{t\bar{t}}\left(1+ a_L \fl + a_R \fr\right)  
, \nonumber \\
A_{FB} &=& A_{FB}^{ 0} \left(1+ b_L \fl + b_R \fr \right). 
\eea 
Here $\sigma^{0}_{t\bar{t}}$ and $A_{FB}^{0}$ are the cross section of top quark pair and the forward and backward asymmetry of top quarks in the SM, respectively. The coefficients $a_{L/R}$ and $b_{L/R}$ describe the interference effects between the SM and anomalous couplings.  The detailed expressions of $a_{L/R}$ and $b_{L/R}$ are given in the Appendix. It is straightforward to determine $\mathcal{F}_L$  and $\mathcal{F}_R$ from $\sigma_{t\bar{t}}$ and $A_{FB}$ as follows:
\bea
\mathcal{F}_L&=\dfrac{b_R}{a_Lb_R-a_Rb_L}\left(\dfrac{\Delta\sigma_{t\bar{t}}}{\sigma_{t\bar{t}}^0}-\dfrac{a_R}{b_R}\dfrac{\Delta A_{FB}}{A_{FB}^0}\right),\label{eq:FL}\\
\mathcal{F}_R&=\dfrac{-b_L}{a_Lb_R-a_Rb_L}\left(\dfrac{\Delta\sigma_{t\bar{t}}}{\sigma_{t\bar{t}}^0}-\dfrac{a_L}{b_L}\dfrac{\Delta A_{FB}}{A_{FB}^0}\right),
\label{eq:FR}
\eea
where $\Delta\sigma_{t\bar{t}}\equiv(\sigma_{t\bar{t}}-\sigma_{t\bar{t}}^{0})$ and $\Delta A_{FB}\equiv(A_{FB}-A_{FB}^{0})$. 

\begin{figure}
\includegraphics[width=0.32\textwidth]{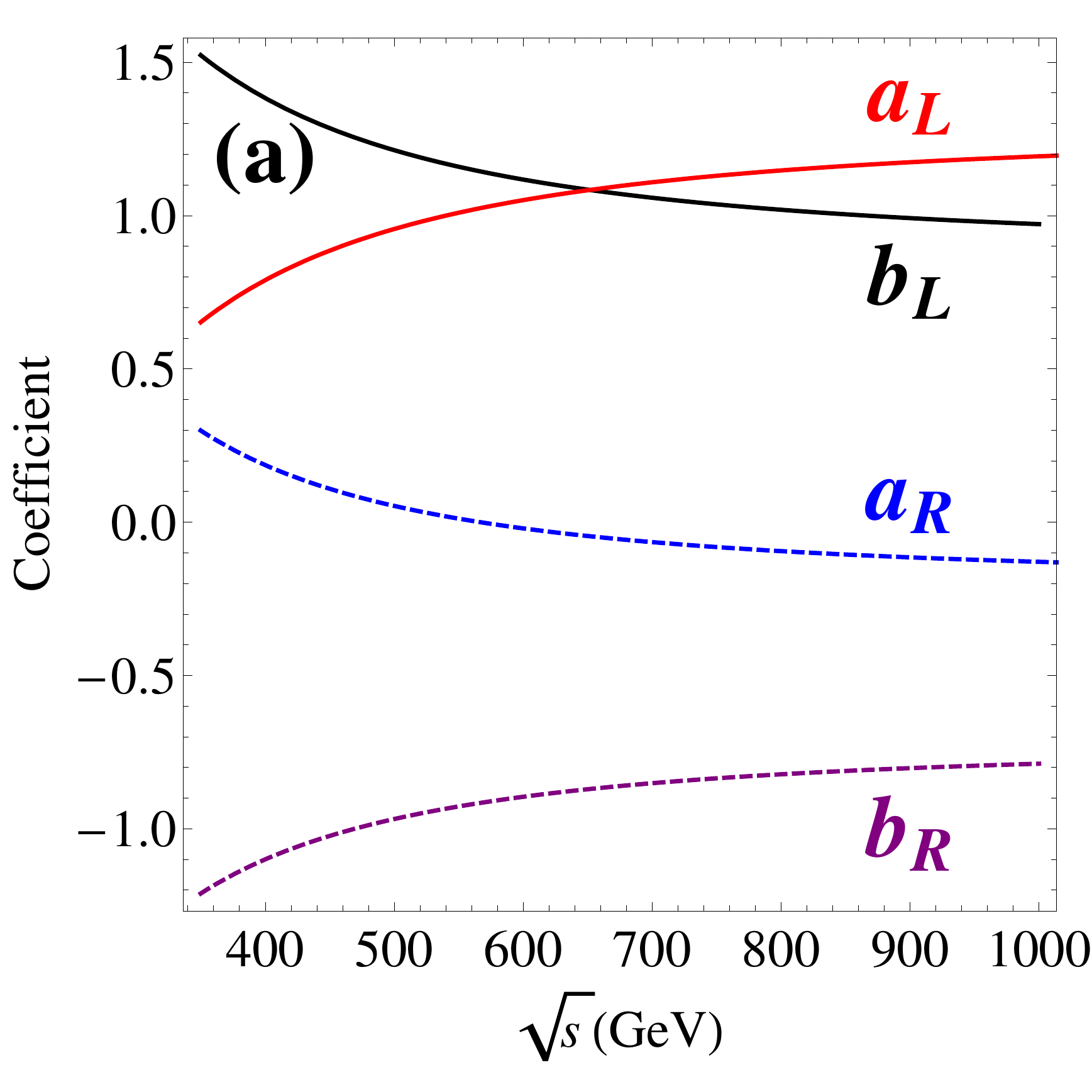}
\includegraphics[width=0.32\textwidth]{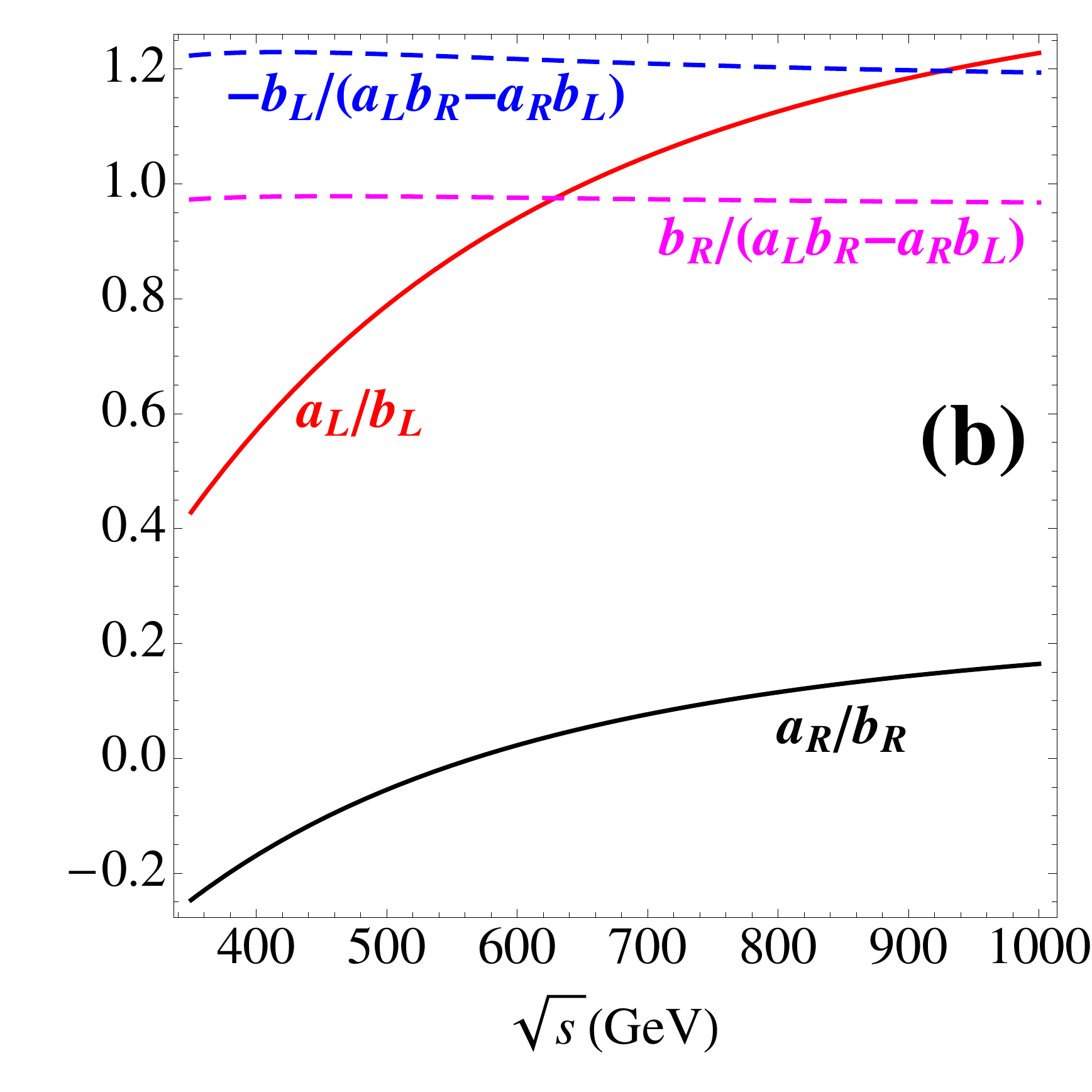}
\caption{\it Dependence on the collider energy $\sqrt{s}$: (a) the coefficients and (b) the ratios. 
}
\label{Fig:coeff}
\end{figure}

The values of the coefficients of $a_{L}$, $a_R$, $b_L$ and $b_R$ depend on the collider energy ($\sqrt{s}$). 
Figure~\ref{Fig:coeff}(a) displays those coefficients as  a function of the collider energy $\sqrt{s}$. Various ratios of those coefficients are also plotted in Fig.~\ref{Fig:coeff}(b), which shows $b_R/(a_L b_R - a_R b_L) \sim 0.97$ and $-b_L/(a_L b_R-a_R b_L) \sim 1.21$~.  As a result, Eqs.~\ref{eq:FL} and \ref{eq:FR} can be approximated as follows:
\bea
\mathcal{F}_L &\sim& 0.97 \left(\dfrac{\Delta\sigma_{t\bar{t}}}{\sigma_{t\bar{t}}^0}-\dfrac{a_R}{b_R}\dfrac{\Delta A_{FB}}{A_{FB}^0}\right) \sim 0.97\dfrac{\Delta\sigma_{t\bar{t}}}{\sigma_{t\bar{t}}^0}, \\
\mathcal{F}_R&\sim&1.21 \left(\dfrac{\Delta\sigma_{t\bar{t}}}{\sigma_{t\bar{t}}^0}-\dfrac{a_L}{b_L}\dfrac{\Delta A_{FB}}{A_{FB}^0}\right),
\eea
where we ignore the $a_R/b_R$ term in the second step as $|a_R/b_R| \lesssim 0.2$ in the region of $400~{\rm GeV} \leq \sqrt{s}\leq 1000~{\rm GeV}$. The ratio $a_L/b_L$ varies from 0.4 to 1.2 in the same energy regime. Note that the above approximation serves only as a guide line for understanding the dependence of $\mathcal{F}_{L/R}$ on $\sigma_{t\bar{t}}$ and $A_{FB}$. In practice one has to combine the measurements of both $\sigma_{t\bar{t}}$ and $A_{FB}$ in order to determine $\mathcal{F}_L$ accurately. 

\begin{figure}
\includegraphics[width=0.32\textwidth]{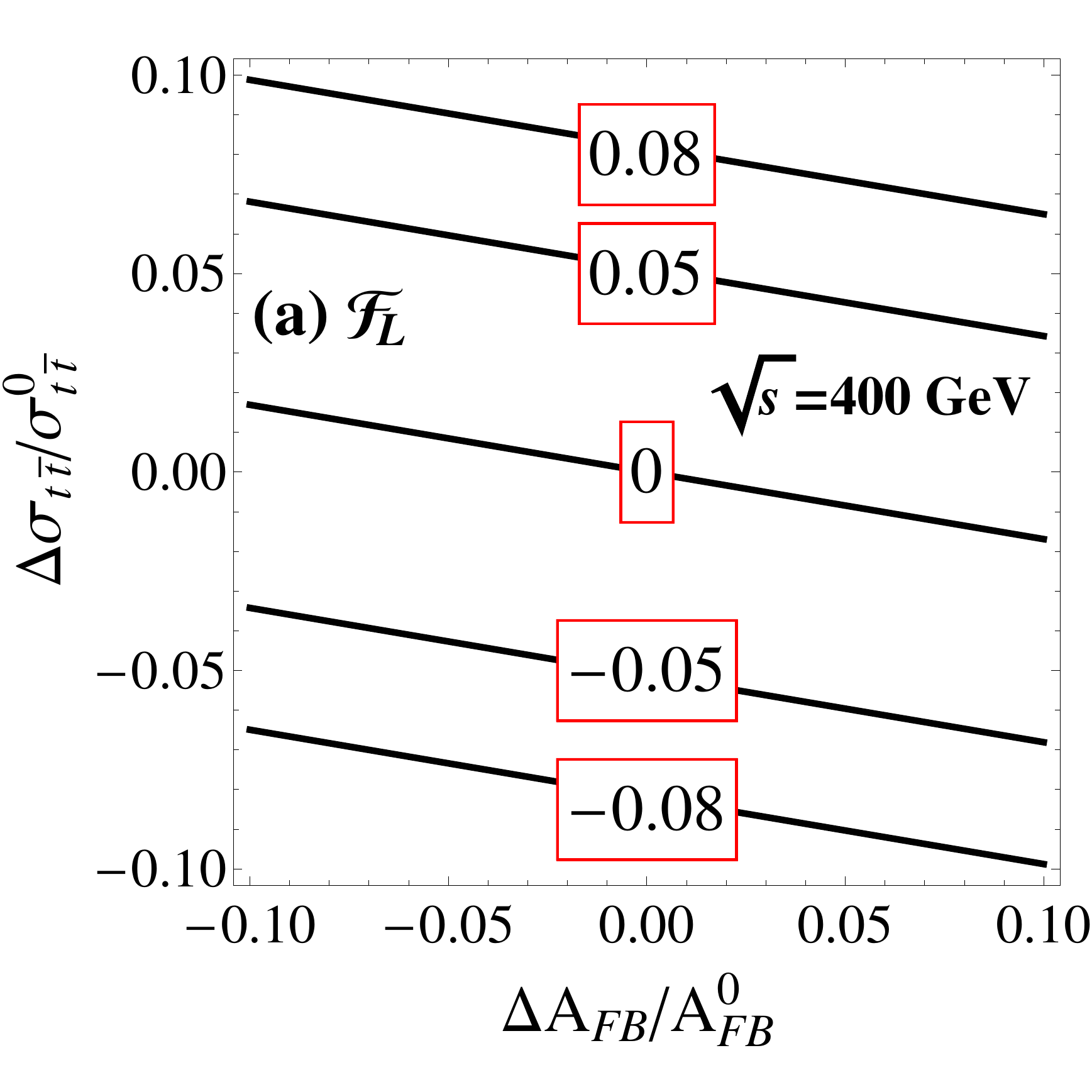}
\includegraphics[width=0.32\textwidth]{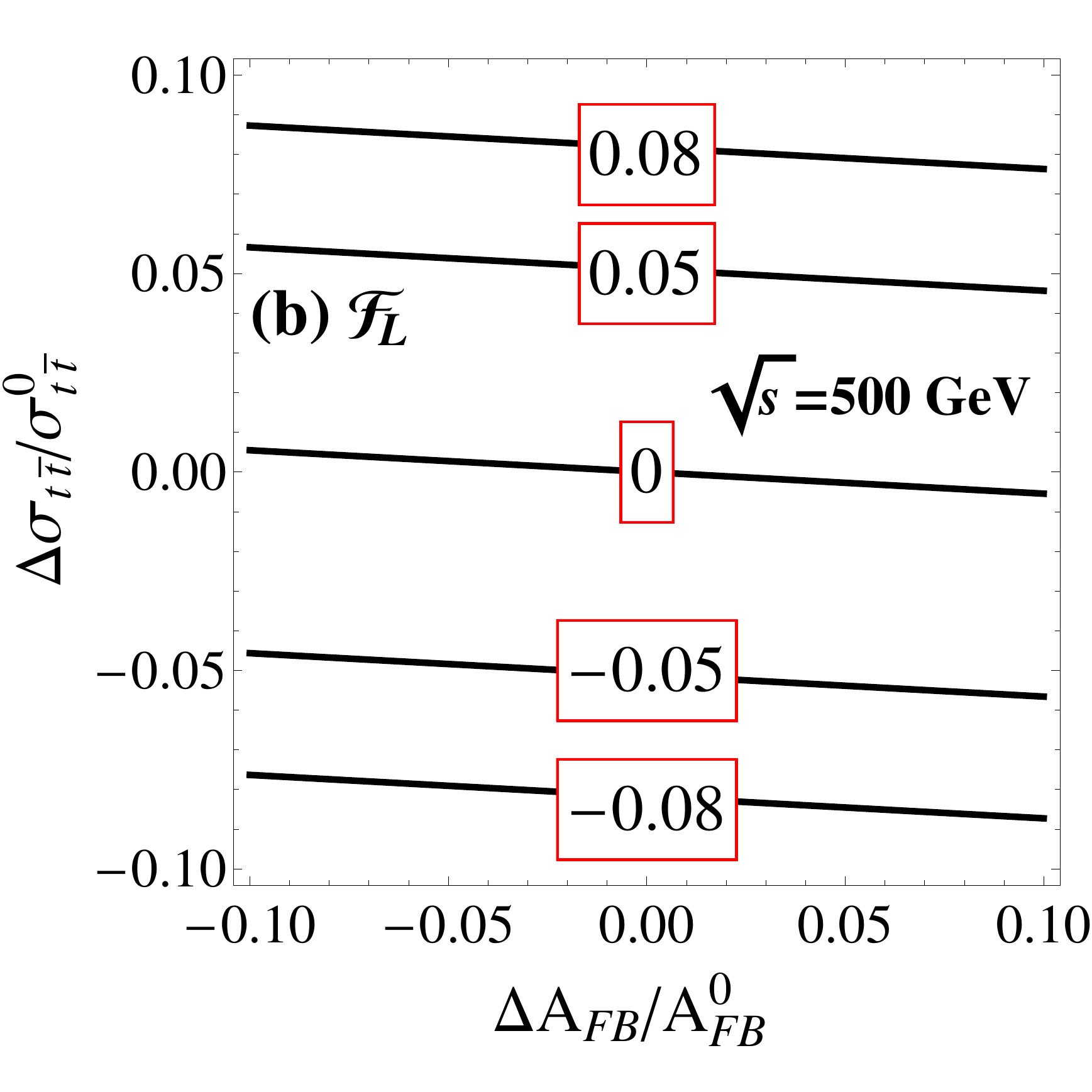}
\includegraphics[width=0.32\textwidth]{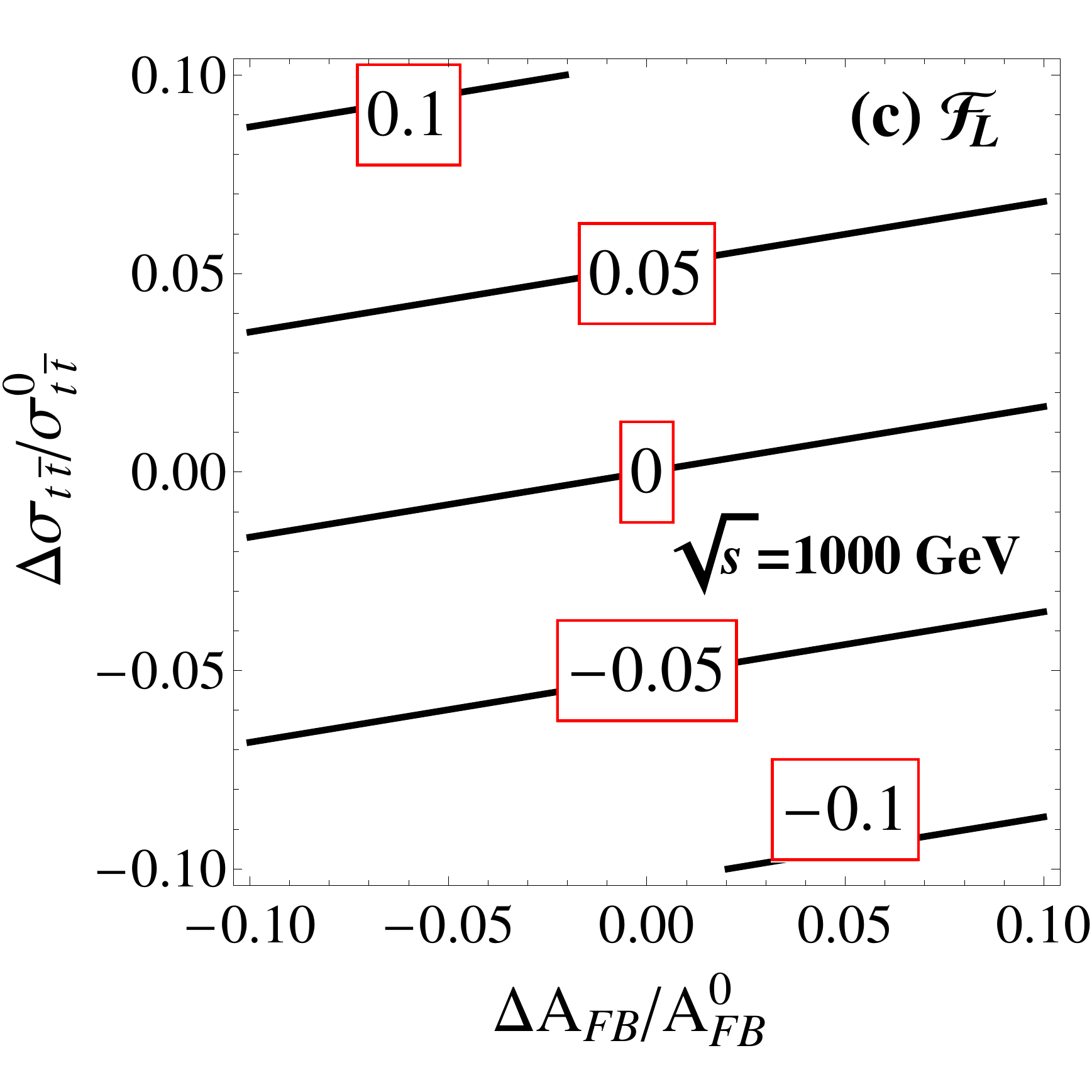}\\
\includegraphics[width=0.32\textwidth]{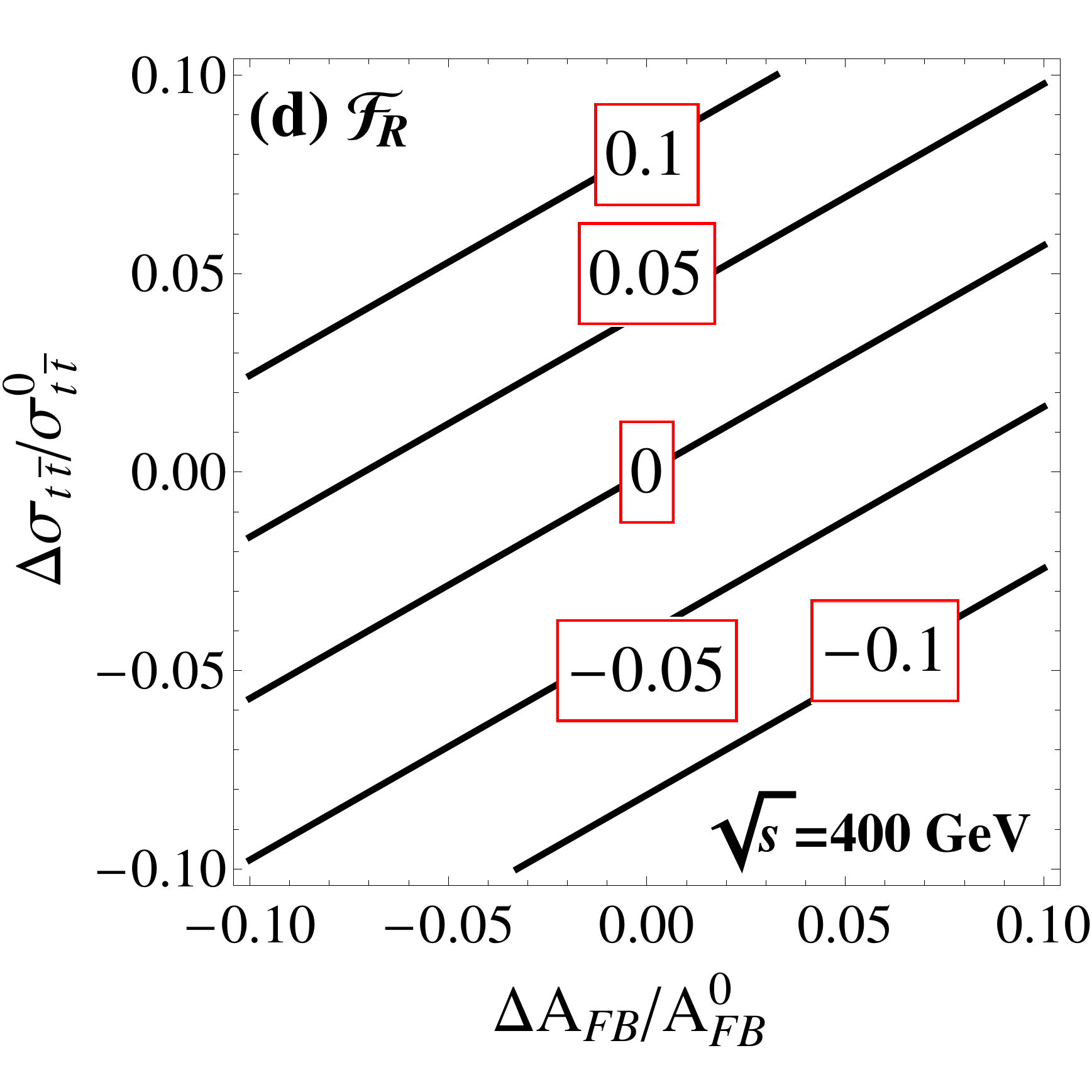}
\includegraphics[width=0.32\textwidth]{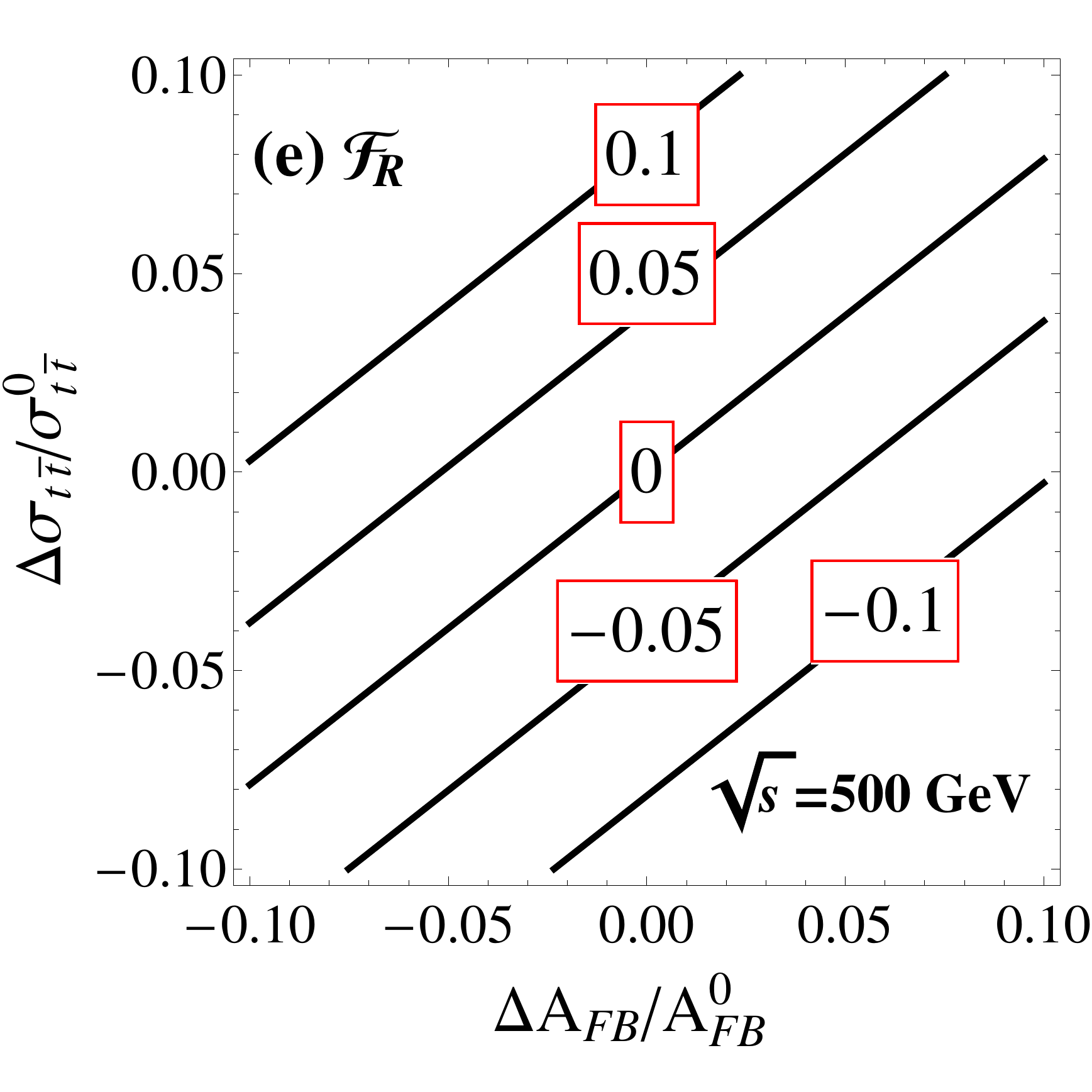}
\includegraphics[width=0.32\textwidth]{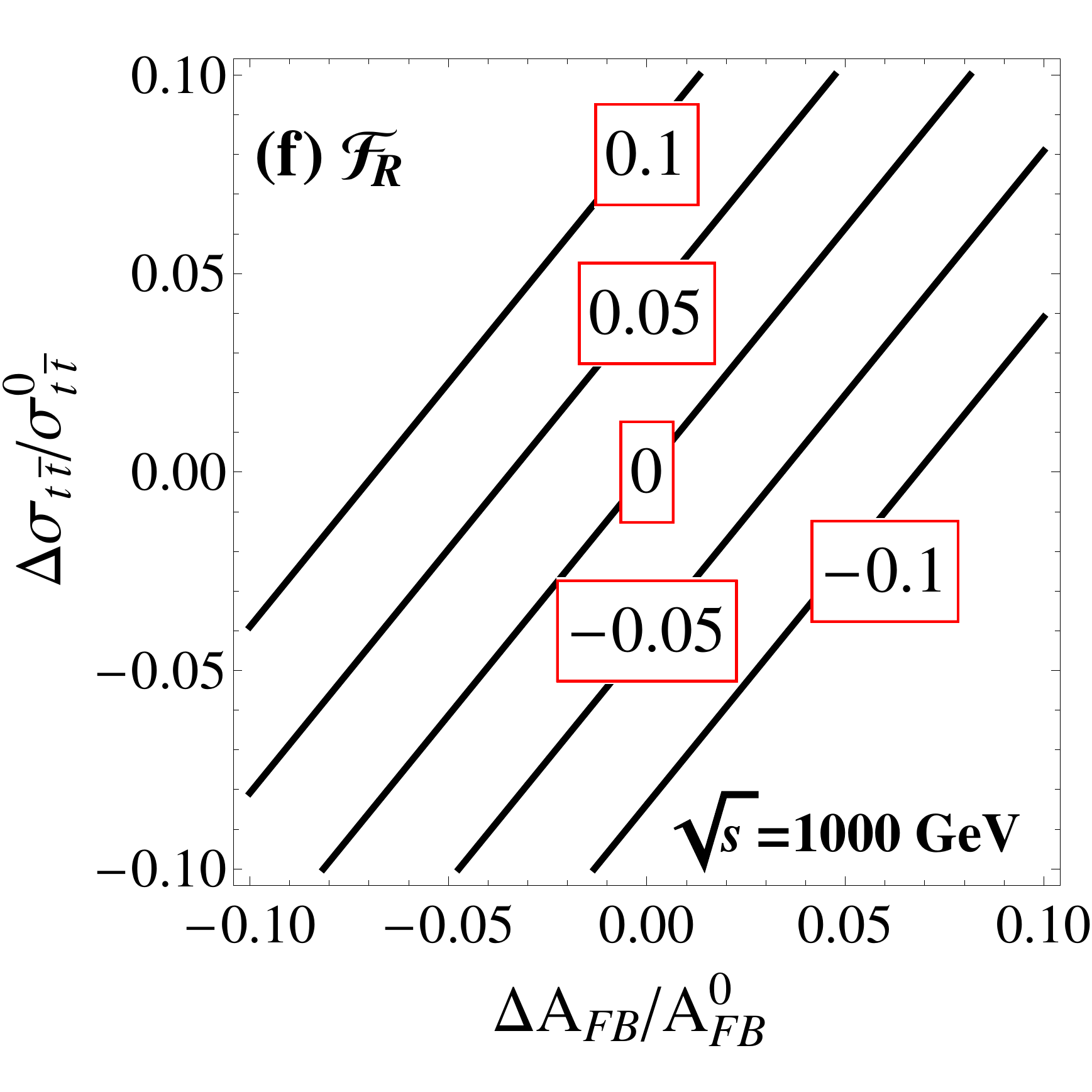}
\caption{\it The contours of $\mathcal{F}_{L/R}$ in the plane of $\Delta\sigma_{t\bar{t}}/\sigma_{t\bar{t}}^0$ and $\Delta A_{FB}/A_{FB}^0$ for $\sqrt{s}=$400, 500 and 1000~GeV. }
\label{Fig:FLR}
\end{figure}

Obviously, $\mathcal{F}_L$ depends mainly on the cross section measurement. We plot in Fig.~\ref{Fig:FLR} the contour of $\mathcal{F}_R$ in the plane of $\Delta \sigma_{t\bar{t}}/\sigma^0_{t\bar{t}}$ and $\Delta A_{FB}/A^0_{FB}$  for three collider energies: (a) $\sqrt{s}=400~{\rm GeV}$, (b) 500~GeV and (c) 1000~GeV.  The slope of the contour lines is determined only by the ratio $a_R/b_R$; see Eq.~\ref{eq:FL}. The difference of $\mathcal{F}_L$ contour lines at $\sqrt{s}=400~{\rm GeV}$ and 1000~GeV can be easily understood from the $a_R/b_R$ curve shown in Fig.~\ref{Fig:coeff}(b), which shows $a_R/b_R<0$ for $\sqrt{s}< 566~{\rm GeV}$ and $a_R/b_R>0$ for $\sqrt{s}>566~{\rm GeV}$. 

The determination of $\mathcal{F}_R$ relies on both $\sigma_{t\bar{t}}$ and $A_{FB}$ measurements.  
We also plot the contours of $\mathcal{F}_{L/R}$ in the plane of  $\Delta\sigma_{t\bar{t}}/\sigma_{t\bar{t}}^{0}$ and $\Delta A_{FB}/A_{FB}^{0}$ in Fig.~\ref{Fig:FLR} (d, e, f) for the three collider energies. Again, the slope of $\mathcal{F}_{R}$ contour lines depends on the ratio of $a_L/b_L$: see Eq.~\ref{eq:FR}. Since the ratio $a_L/b_L$ is always positive for $400~{\rm GeV}\leq \sqrt{s}\leq 1000~{\rm GeV}$, the $\mathcal{F}_R$ contours are very similar  for the three collider energies.

\begin{figure}
\includegraphics[width=0.32\textwidth]{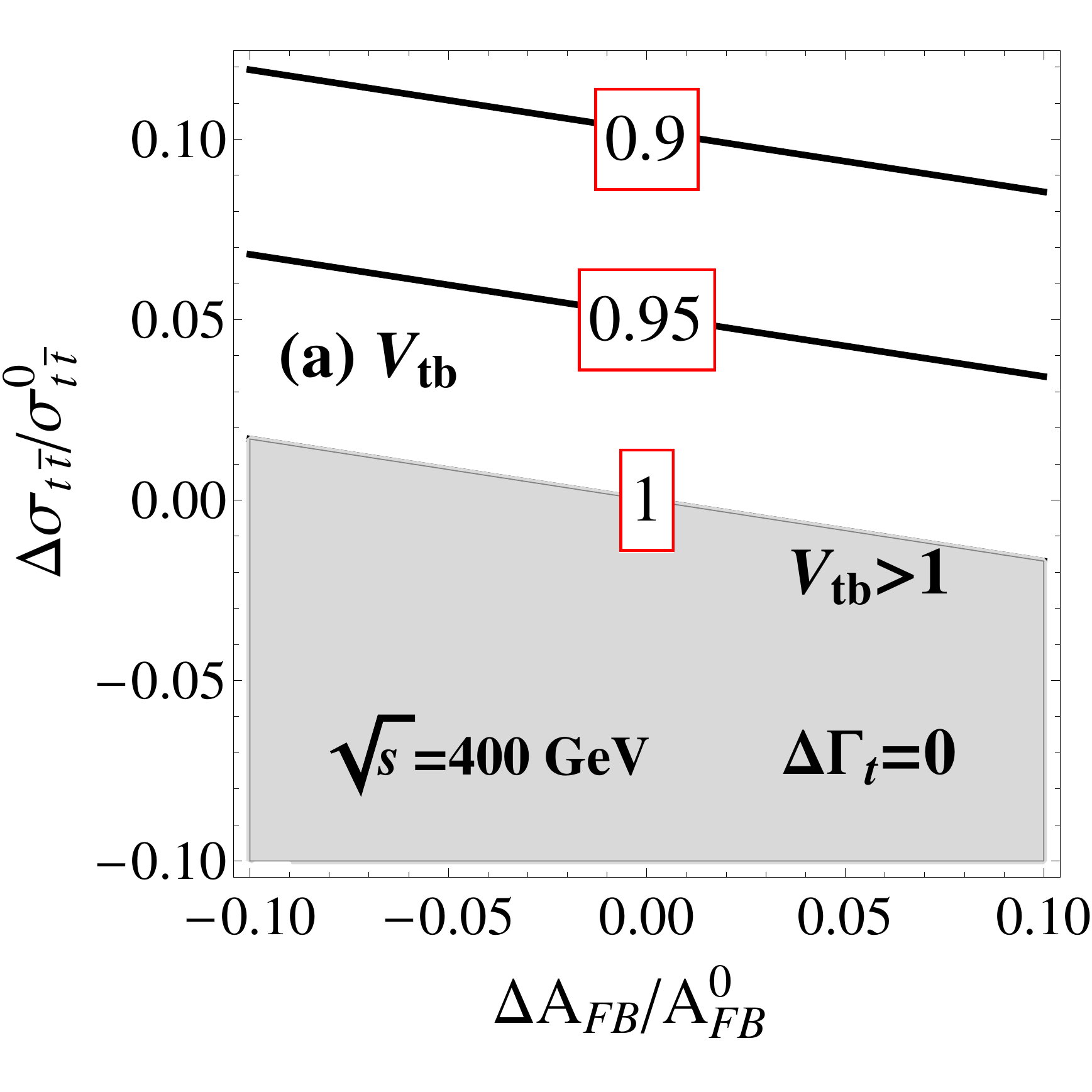}
\includegraphics[width=0.32\textwidth]{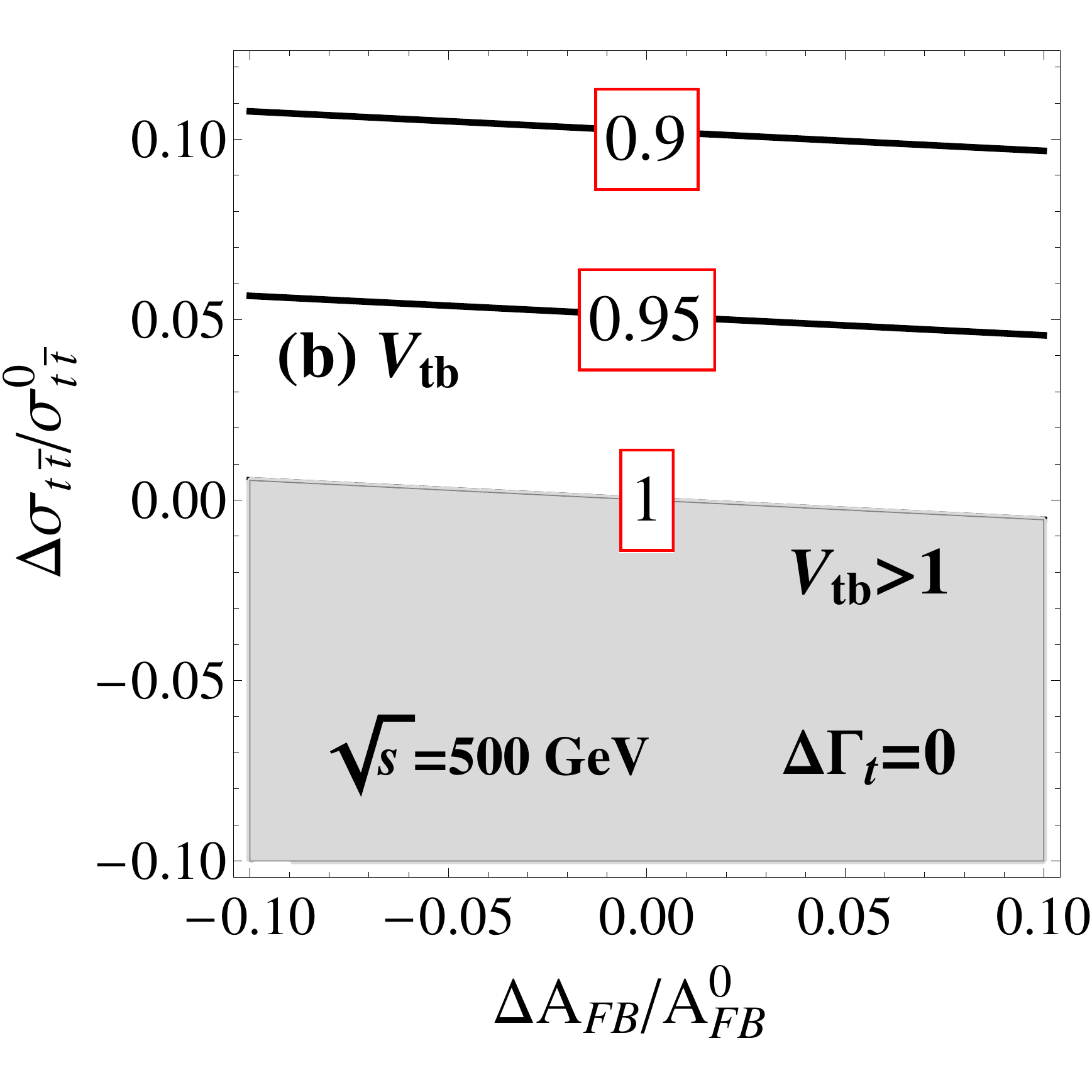}
\includegraphics[width=0.32\textwidth]{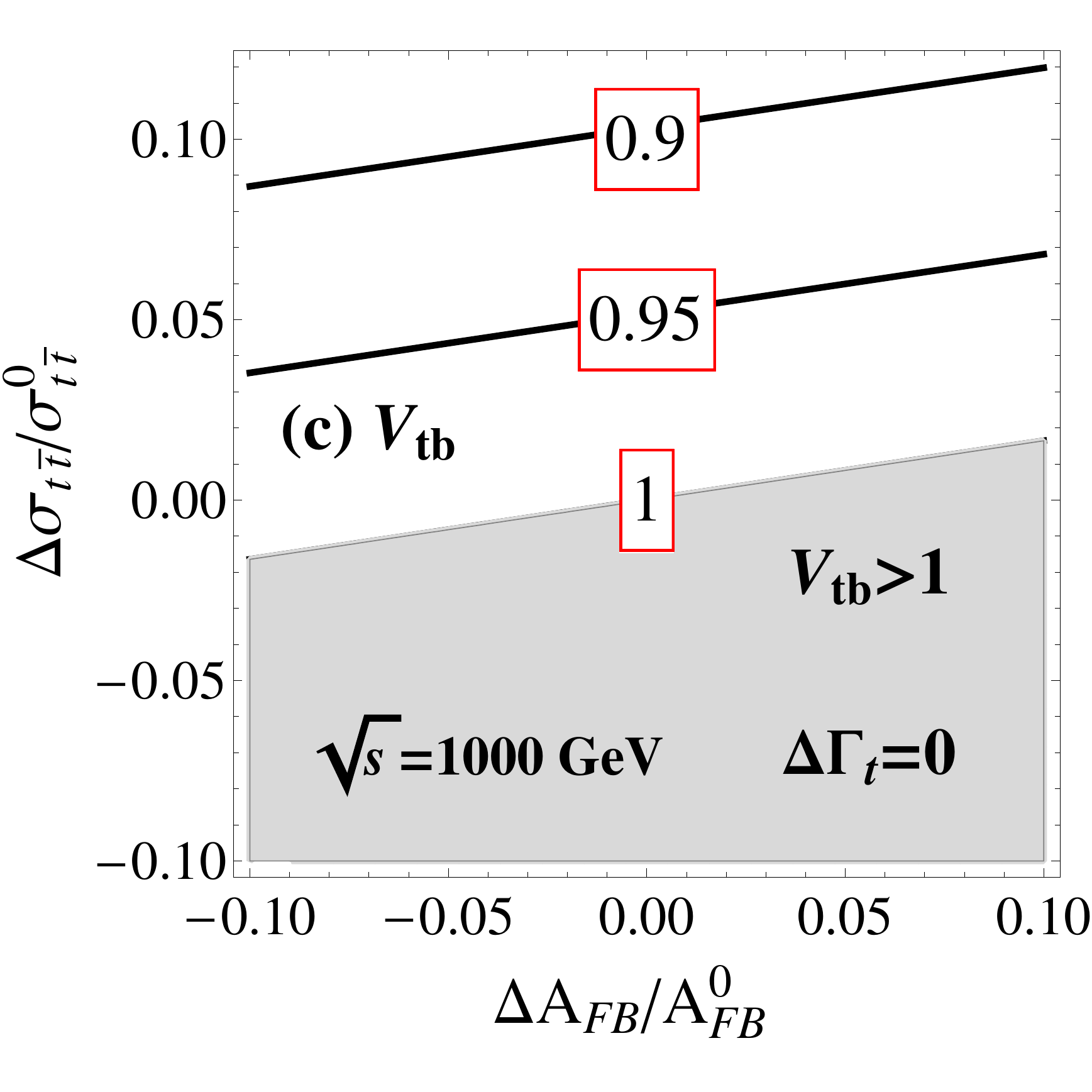}
\caption{\it The contours of $V_{tb}$ in the plane of $\Delta\sigma_{t\bar{t}}/\sigma_{t\bar{t}}^0$ and $\Delta A_{FB}/A_{FB}^0$ for $\sqrt{s}=$400, 500 and 1000~GeV under the assumption of  $\Delta\Gamma_t\simeq 0$. The gray shaded region represents the parameter space of $V_{tb}>1$. }
\label{Fig:vtb}
\end{figure}

There is a strong anti-correlation between $\Delta V_{tb}$ and $\Delta \sigma_{t\bar{t}}$ as 
\bea
\Delta V_{tb}\approx \frac{1}{2}\frac{\Delta\Gamma_t}{\Gamma_t^{0}}-0.97\dfrac{\Delta\sigma_{t\bar{t}}}{\sigma_{t\bar{t}}^0}.
\eea 
For simplicity we assume $\Delta\Gamma_t\simeq 0$, i.e. the top-quark width is exactly the same as the SM theory prediction. Figure~\ref{Fig:vtb} displays the contour of $V_{tb}\equiv 1+\Delta V_{tb}$ in the plane of $\Delta \sigma_{t\bar{t}}/\sigma^0_{t\bar{t}}$ and $\Delta A_{FB}/A^0_{FB}$. The shaded region represents $V_{tb}>1$ (i.e. $\Delta V_{tb} > 0$) which violates the unitarity condition. Demanding $V_{tb} \leq 1$ in NP models implies that $\sigma_{t\bar{t}}$ would be likely enhanced. In order to precisely determine the value of $V_{tb}$ matrix element, both measurements of $\sigma_{t\bar{t}}$ and  $A_{FB}$  are needed. However,  we emphasize that,  at an {\it unpolarized} $e^+e^-$ collider with $\sqrt{s}= 500~{\rm GeV}$, the measurement of $\sigma_{t\bar{t}}$ alone is already good to probe $\mathcal{F}_L$ which can be used to determine $\Delta V_{tb}$. For example, a 5\% deviation in the $\sigma_{t\bar{t}}$ cross section indicates $V_{tb}\simeq 0.95$, regardless of the $A_{FB}$ measurement.

\subsection{Error analysis}

Next we discuss the uncertainties of extracting $\mathcal{F}_L$ and $\mathcal{F}_R$ out of the cross section and asymmetry measurements. 
Based on the error propagation equation of the weighted-sum functions, the errors of $\mathcal{F}_{L}$ and $\mathcal{F}_{R}$ are,
\bea
\delta\mathcal{F}_L &=& \left|\dfrac{b_R}{a_Lb_R-a_Rb_L}\right|\sqrt{\left(\dfrac{\delta\sigma_{t\bar{t}}}{\sigma_{t\bar{t}}^0}\right)^2+\left(\dfrac{a_R}{b_R}\right)^2\left(\dfrac{\delta A_{FB}}{A_{FB}^0}\right)^2},\nonumber\\
\delta\mathcal{F}_R &=& \left|\dfrac{b_L}{a_Lb_R-a_Rb_L}\right|\sqrt{\left(\dfrac{\delta\sigma_{t\bar{t}}}{\sigma_{t\bar{t}}^0}\right)^2+ \left(\dfrac{a_L}{b_L}\right)^2 \left(\dfrac{\delta A_{FB}}{A_{FB}^0}\right)^2},
\eea
where $\delta\sigma_{t\bar{t}}$ and $\delta A_{FB}$ denote the total errors of the $\sigma_{t\bar{t}}$ and $A_{FB}$ defined as follows:
\begin{align}
\delta\sigma_{t\bar{t}}&\equiv\sqrt{(\delta\sigma_{t\bar{t}})_{sys.}^2+(\delta\sigma_{t\bar{t}})_{stat.}^2},&
\delta A_{FB}&\equiv\sqrt{(\delta A_{FB})_{sys.}^2+(\delta A_{FB})_{stat.}^2}.
\end{align}
The statistical errors of  $\sigma_{t\bar{t}}$ and $A_{FB}$, which are normalized to the SM predictions, are
\bea
(\delta\sigma_{t\bar{t}}/\sigma_{t\bar{t}}^0)_{stat.}&=&\sqrt{1/(\mathcal{L}\sigma_{t\bar{t}}^0)}~, \nn\\
(\delta A_{FB}/A_{FB}^0)_{stat.}&=&\sqrt{(1-(A_{FB}^{0})^2)/(\mathcal{L}\sigma_{t\bar{t}}^0)}.
\eea
For an integrated luminosity of $500~{\rm fb}^{-1}$ and collider energy $\sqrt{s}=500~{\rm GeV}$, $(\delta \sigma_{t\bar{t}}/\sigma_{t\bar{t}}^0)_{stat.}\simeq (\delta A_{FB}/A_{FB}^0)_{stat.}\sim 0.002$.

The systematic uncertainty arises from a lot of experimental effects, e.g. cut acceptance, $b$-tagging efficiency, detector resolution, luminosity or different hadronization of  $t\bar{t}$ events, etc. Those systematic uncertainties will have to be estimated at a later stage, but they are expected to be small~\cite{Amjad:2013tlv}. The LEP-I reported a systematic uncertainty on $R_b$ of 0.28~\%~\cite{ALEPH:2005ab} which may serve as a guide line for values to be expected at the future $e^+ e^-$ collider. In this work, the systematic error of $\sigma_{t\bar{t}}$ relative to the SM prediction is assumed to be around 1\%, i.e. $(\delta\sigma_{t\bar{t}}/\sigma_{t\bar{t}}^0)_{sys.}=0.01$~\cite{Baer:2013cma,Amjad:2013tlv}. Table~\ref{tbl:error} displays the statistical and systematic errors of $m_t$, $\Gamma_t$, $\sigma_{t\bar{t}}$ and $A_{FB}$ used in this study.

\begin{figure}
\includegraphics[width=0.32\textwidth]{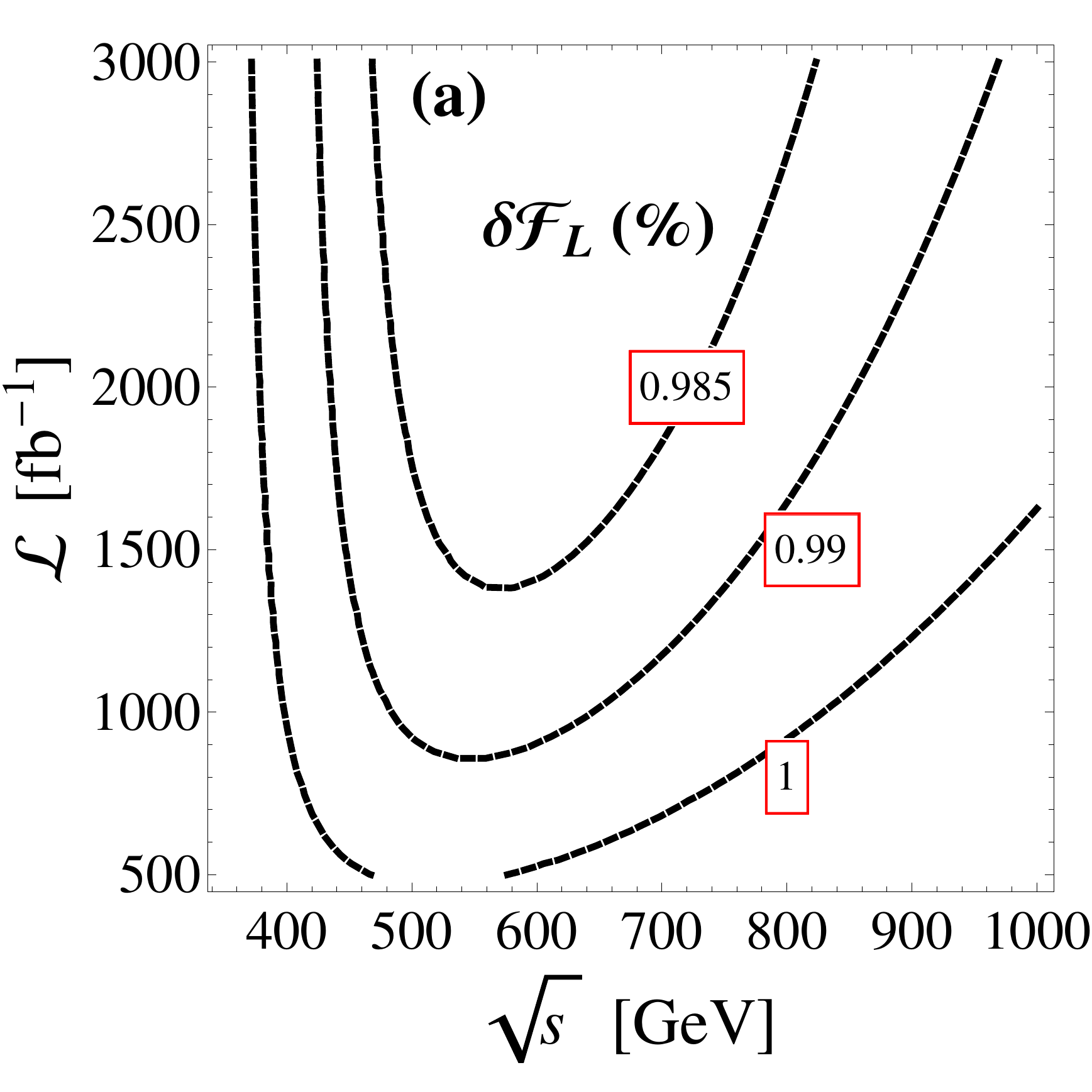}
\includegraphics[width=0.32\textwidth]{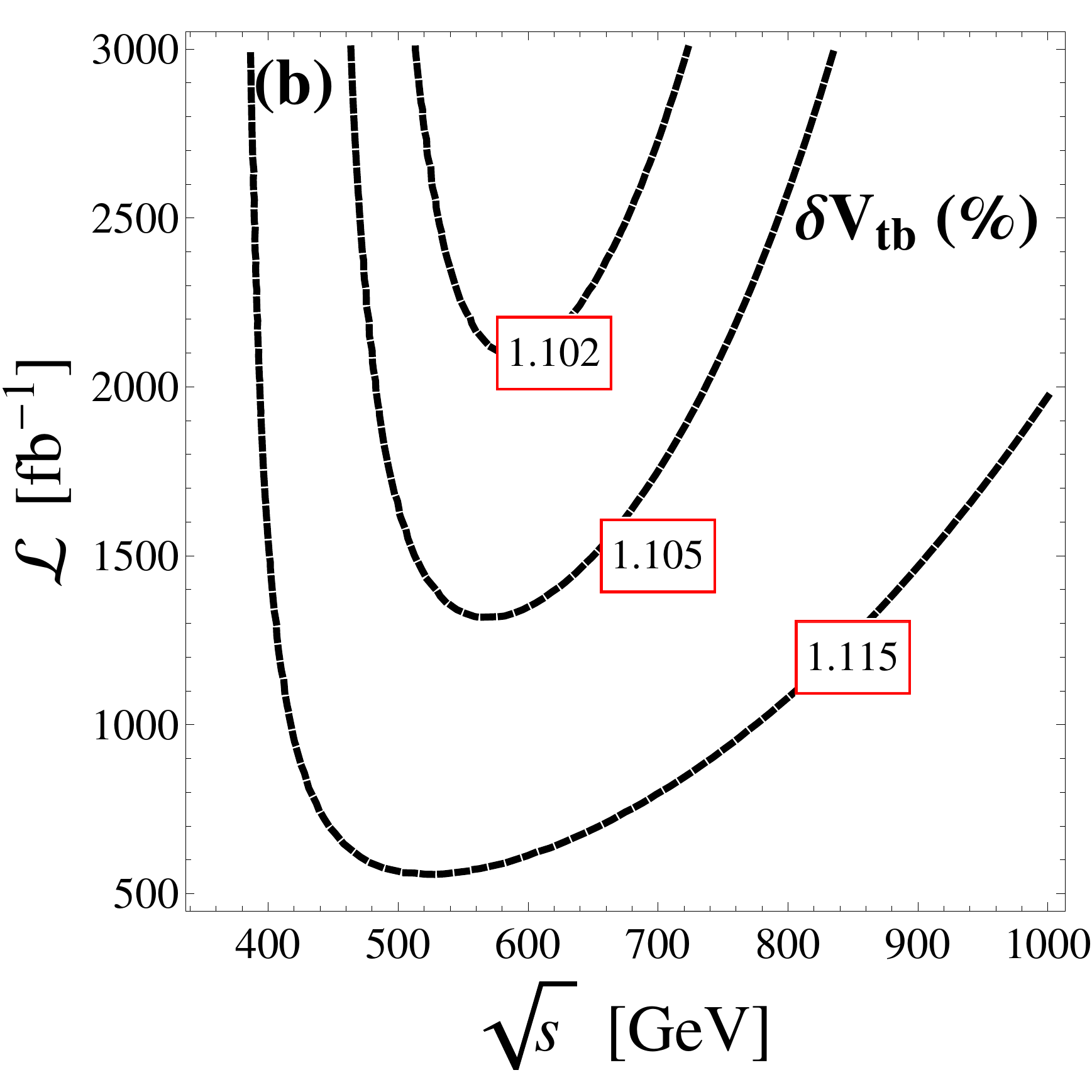}
\caption{\it The contours of uncertainties of $\mathcal{F}_L$ and $V_{tb}$ measurements denoted by $\delta\mathcal{F}_L$ and $\delta V_{tb}$ in the plane of  the collider energy $\sqrt{s}$ (GeV) and integrated luminosity $\mathcal{L}$ ($fb^{-1}$).}
\label{Fig:error}
\end{figure}

Figure~\ref{Fig:error}(a) displays the contours of $\delta\mathcal{F}_L$ in the plane of the collider energy  $\sqrt{s}$ (GeV) and integrated luminosity $\mathcal{L}$ ($\rm{fb}^{-1}$). It shows that $\mathcal{F}_L$ can be measured with an accuracy of percentage, e.g. $\delta\mathcal{F}_L\leq 1\%$. The uncertainty,  which is dominated by the systematic error, cannot be further improved by increasing the collider energy and accumulating more luminosities. One then can translate the uncertainty of $\mathcal{F}_L$ measurement to the $ V_{tb}$ measurement as following 
\bea
\delta V_{tb}=\sqrt{\frac{1}{4}\left(\dfrac{\delta\Gamma_t}{\Gamma_t^{0}}\right)^2+(\delta F_L)^2}~.
\eea
Figure~\ref{Fig:error}(b) displays that $\delta V_{tb}$ can be measured as accurately as $\delta\mathcal{F}_L$, e.g. $\delta V_{tb}\leq 1.1\%$, and it is not sensitive to the collider energy or the integrated luminosities. Therefore, we argue that it is enough to determine $V_{tb}$  at the $e^+e^-$ collider with $\sqrt{s}\gsim 400$ GeV. We also note that the cross section measurement alone is adequate to determine $\mathcal{F}_L$ when $\sqrt{s}\geq 500~{\rm GeV}$. To further constrain the  $\delta V_{tb}$, it is necessary to reduce the systematic error and improve the measurement of top quark width.

\begin{table}
\caption{The statistical and systematic uncertainties of $m_t$, $\Gamma_t$, $\sigma_{t\bar{t}}$ and $A_{FB}$. The statistic uncertainties of $m_t$ and $\Gamma_t$ are quoted from Ref.~\cite{Gomez-Ceballos:2013zzn} while the statistical error of $\sigma_{t\bar{t}}$ and $A_{FB}$ are obtained at a 500~GeV collider with an integrated luminosity of $500{~\rm fb}^{-1}$. The systematic error of $\Gamma_t$, $\sigma_{t\bar{t}}$ and $A_{FB}$ are assumed to be $\sim 1\%$ throughout this work. The uncertainties of $R_b$ measured at LEP-I are listed for reference~\cite{ALEPH:2005ab}. 
}
\begin{tabular}{c|cccc|c}
\hline
     &    $m_t$   &   $\Gamma_t$  & $\sigma_{t\bar{t}}$ & $A_{FB}$  & $R_b$ (LEP-I)\\
\hline
stat. &  0.006\%  &    0.5\%      &       0.2\%       &  0.2\%  &   0.44\% \\
\hline
sys.  &  -        &    1\%     &        1\%          &  1\%   &  0.28\%\\
\hline
\end{tabular}
\label{tbl:error}
\end{table}

\section{Implications on new physics models}

For illustration we examine the impact of the $V_{tb}$ and $\mathcal{F}_{L,R}$ measurements on NP models. 
We begin with the so-called fourth-generation model~\cite{Bose:1979vz,Gronau:1984dn,Botella:1985gb,Alwall:2006bx,Kribs:2007nz,Kuflik:2012ai,Eberhardt:2012gv,Eberhardt:2012ck}. The perturbative fourth generation is disfavored as it would induce a large gluon-gluon-Higgs effective coupling which produces a too large cross section of the Higgs boson production to obey the current data~\cite{Kuflik:2012ai,Eberhardt:2012gv,Eberhardt:2012ck}. However, vector-like quarks, which exhibit the decoupling behavior, would not affect the Higgs production too much if the vector-like quarks are very heavy. The vector-quarks might modify the CKM matrix elements, depending on their quantum number. It is critical to directly measure the $V_{tb}$ element which is complementary to the $H$-$g$-$g$ coupling measurement.  
Next we use the fourth-generation model as a good example to discuss the impact of $V_{tb}$ measurement. Our results can be extended easily to NP models with extra heavy quarks which modify $V_{tb}$ through the mixing of the heavy quarks and the third generation quarks. 

Neglecting the possible CP-violating phases beyond $3\times 3$ CKM matrix $V_{3\times 3}$, we can parametrize the $4\times 4$ unitary matrix $V_{4\times 4}$ with  $V_{3\times 3}$ and extra mixing angles as follows~\cite{Alwall:2006bx},
\bea
V_{4\times 4}=R_{34}(\theta_{34})R_{24}(\theta_{24})R_{14}(\theta_{14})
\begin{pmatrix}
V_{3\times 3}^0 & \textbf{1}_{1\times 3} \\
\textbf{1}_{3\times 1} & 1
\end{pmatrix},
\eea
where $V_{3\times 3}^0$ denote the CKM matrix involving the three generation fermions in the SM, 
$R_{ij}(\theta_{ij})$ is the rotation matrix in the $(i,j)$ flavour plane with rotation angle $\theta_{ij}$. Since $V_{tb}^0 \gg V^0_{cb,ub}$ in $V_{3\times 3}^0$, one can approximate the $V_{tb}$ matrix element as
\bea
V_{tb} \equiv V_{tb}^0+\Delta V_{tb}\approx \cos\theta_{34}V_{tb}^0 =\cos\theta_{34}.
\eea
Thus, the dependence of $\cos\theta_{34}$ on the $\sigma_{t\bar{t}}$ and $A_{FB}$ measurements is exactly the same as those of $V_{tb}$ shown in Fig.~\ref{Fig:vtb}.  The uncertainties on $\cos\theta_{34}$ are also identical to those in Fig.~\ref{Fig:error}(b).

\begin{figure}
\includegraphics[width=0.4\textwidth]{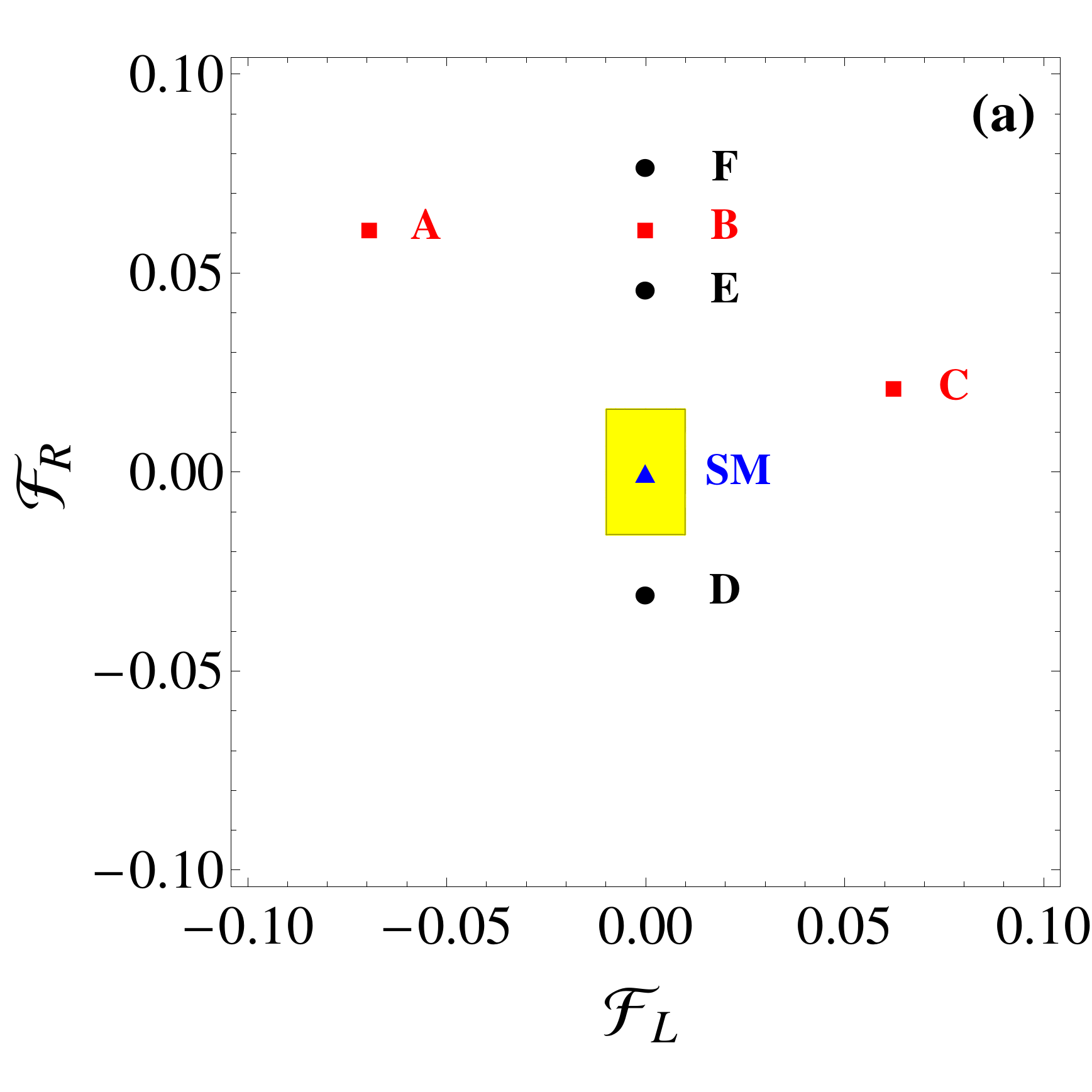}
\includegraphics[width=0.4\textwidth]{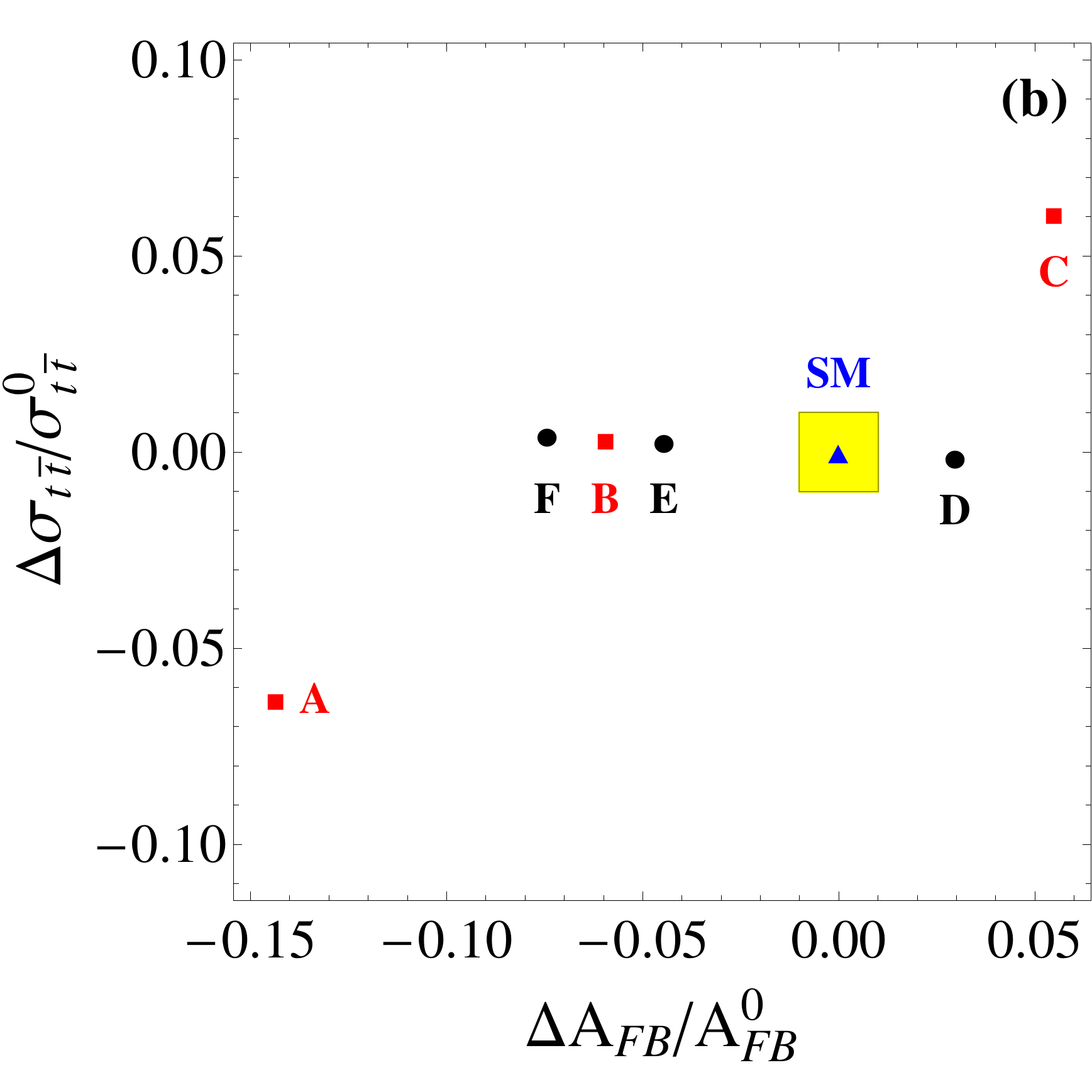}
\caption{\it Projected 68.3\% C.L. bounds (shaded yellow region) on (a) the anomalous $Zt\bar{t}$ copulings and (b) $\sigma_{t\bar{t}}$ and $A_{FB}$ measurements at a 500~GeV $e^+e^-$ collider  with an integrated luminosity of $500~{\rm fb}^{-1}$. Several benchmark points of NP models are also shown: the red box denotes the extra dimensional models: A (Gherghetta et al~\cite{Cui:2010ds}), B (Carena et al~\cite{Carena:2006bn}), C (Hostanoi et al~\cite{Hosotani:2005nz}), the black disk denotes the composite models: D (Grojean et al~\cite{Grojean:2013qca}), E (Little Higgs ~\cite{Berger:2005ht}), F (Pomarol et al~\cite{Pomarol:2008bh}).  }
\label{Fig:zttNP}
\end{figure}

A by-product of measuring $V_{tb}$ in the process of $e^+ e^- \to t\bar{t}$ is to determine both $\fl$ and $\fr$ precisely, which can be used to distinguish different NP models. The $Zt\bar{t}$ couplings could be modified in various NP models, e.g. the extra dimension models~\cite{Cui:2010ds,Carena:2006bn,Hosotani:2005nz}, composite models~\cite{Grojean:2013qca,Berger:2005ht,Pomarol:2008bh} and non-Abelian extension models~\cite{Hsieh:2010zr,Cao:2012ng,Cao:2015lia}. Several benchmark points of those NP models mentioned above are nicely summarized in Table I of Ref.~\cite{Richard:2014upa}. Figure~\ref{Fig:zttNP}(a) displays the expected precision of the $\mathcal{F}_L$ and $\mathcal{F}_{R}$ measurements and the benchmark points of NP models at an unpolarized $e^+e^-$ collider with $\sqrt{s}=500~{\rm GeV}$ and $\mathcal{L}=500~{\rm fb}^{-1}$.  The shaded region denotes the expected uncertainties based on the error analysis discussed above. The triangle symbol represents the SM, the black disk denotes the  composite model while the red box the extra dimension models. 
For $\sqrt{s}=500~{\rm GeV}$, the $\mathcal{F}_{L}$ and $\mathcal{F}_R$ anomalous couplings are related to $\sigma_{t\bar{t}}$ and $A_{FB}$  as follows:
\beq
\mathcal{F}_L  =  0.98 \dfrac{\Delta\sigma_{t\bar{t}}}{\sigma_{t\bar{t}}^0}-0.05\dfrac{\Delta A_{FB}}{A_{FB}^0},~~
\mathcal{F}_R = 1.23 \dfrac{\Delta\sigma_{t\bar{t}}}{\sigma_{t\bar{t}}^0}-0.97 \dfrac{\Delta A_{FB}}{A_{FB}^0}.\nn
\eeq
Figure~\ref{Fig:zttNP}(b) displays the NP models  in the plane of $\Delta \sigma_{t\bar{t}}/\sigma_{t\bar{t}}^0$ and $\Delta A_{FB}/A^0_{FB}$.  Those NP models can be easily discriminated if they modify the $Zt\bar{t}$ anomalous couplings sizably.

\section{Summary}

In this work we proposed to measure the $V_{tb}$ element of the CKM matrix in the process of $e^+e^- \to t\bar{t}$ without assuming the $3\times 3 $ unitarity of CKM matrix and universality of weak gauge couplings. Four experimental observables are considered in our analysis: the top-quark mass and width, the cross section of top-quark pair production $\sigma_{t\bar{t}}$, and the Forward-Backward asymmetry of the top-quark $A_{FB}$. We first consider the impact of NP effects on the top-quark mass and width which can be measured very precisely from the threshold energy scan experiments at $e^+e^-$ colliders. The would-be measured top-quark width imposes a strong correlation between the deviation of $V_{tb}$ (denoted by $\Delta V_{tb}$) and the deviation of $W$-$t_L$-$b_L$ coupling (denoted by $\mathcal{F}_L$). In order to determine $V_{tb}$, $\mathcal{F}_L$ must be measured from other sources.  Using an effective Lagrangian approach, we perform a model-independent analysis of the interactions among electroweak gauge bosons and the third generation quarks, i.e. the $Wtb$, $Zt\bar{t}$ and $Zb\bar{b}$ couplings. After one imposes the known experimental constraint on the $Z$-$b_L$-$b_L$ coupling, the electroweak $SU(2)_L \times  U(1)_Y$ symmetry of the SM specifies a pattern of deviations of the $Z$-$t_L$-$t_L$ and $W$-$t_L$-$b_L$ couplings, independent of underlying new physics scenarios. The predicted pattern enables us to infer $\mathcal{F}_L$ from the $Zt\bar{t}$ coupling measurement in the process of $e^+ e^- \to t\bar{t}$ at an unpolarized $e^+e^-$ collider. 

The deviations of the $Zt\bar{t}$ coupling are described by the left-handed coupling $\mathcal{F}_L$ and right-handed coupling $\mathcal{F}_R$, which can be determined from $\sigma_{t\bar{t}}$ and $A_{FB}$. We show that $\mathcal{F}_L$ relies mainly on $\Delta \sigma_{t\bar{t}}$ at an unpolarized $e^+e^-$ collider, especially for $\sqrt{s}\geq 500~{\rm GeV}$. It leads to a strong anti-linear correlation between $\Delta V_{tb}$ and $\Delta \sigma_{t\bar{t}}$, $\Delta V_{tb}\approx 0.5\Delta \Gamma_t/\Gamma_t^0 -0.97\Delta\sigma_{t\bar{t}}/\sigma_{t\bar{t}}^0$, where $\Gamma_t^0$ and $\sigma_{t\bar{t}}^0$ denote the top-quark width and the cross section of top-quark pair production, respectively. If the top-quark width is not modified in NP models, requiring $V_{tb}<1$ (i.e. $\Delta V_{tb}\leq 0$) implies that $\sigma_{t\bar{t}}$ will be inevitably enhanced. We also show that the uncertainty of $V_{tb}$ measurement is dominated by the systematic errors which is assumed to be 1\% in this work. 

On the other hand, $\mathcal{F}_R$ is sensitive to the deviations of both $\sigma_{t\bar{t}}$ and $A_{FB}$. One has to combine $\sigma_{t\bar{t}}$ and $A_{FB}$ to obtain a precise value of $\mathcal{F}_R$. Knowing both $\mathcal{F}_L$ and $\mathcal{F}_R$ is important to distinguish new physics models.

\begin{acknowledgments}
We thank Yan-Dong Liu, C.-P. Yuan and Chen Zhang for helpful discussions. The work is supported in part by the National Science Foundation of China under Grand No. 11275009.
\end{acknowledgments}

\appendix

\section{The $\sigma_{t\bar{t}}$ and $A_{FB}$ at a $e^+e^-$ collider}

The cross section $\sigma_{t\bar{t}}$ and the asymmetry $A_{FB}$ of top-quark pairs in the process  $e^+e^- \to \gamma/Z \to t\bar{t}$ are given as follows:
\bea
\sigma_{t\bar{t}} &=& \sigma^{\rm 0}_{t\bar{t}}\left(1+ a_L \fl + a_R \fr\right)  = \sigma^{\rm 0}_{t\bar{t}}\left(1+ a_V F_V + a_A F_A\right), \nonumber \\
A_{FB} &=& A_{FB}^{\rm 0} \left(1+ b_L \fl + b_R \fr \right) =  A_{FB}^{\rm 0} \left(1+ b_V F_V + b_A F_A \right),
\eea 
where 
\begin{align}
F_V&=(\mathcal{F}_L+\mathcal{F}_R/2)/2, & F_A&=(-\mathcal{F}_L+\mathcal{F}_R/2)/2, \nonumber\\
a_V&=a_L+2a_R,& a_A&=2a_R-a_L,\nonumber\\
b_V&=b_L+2b_R,& b_A&=2b_R-b_L.
\end{align}
The $\sigma_{t\bar{t}}^{\rm 0}$ and $A_{FB}^{\rm 0}$ denote the cross section and asymmetry in the SM, respectively, and are given as follows: 
\begin{align}
\sigma_{t\bar{t}}^{\rm 0}&=\frac{\beta (3-\beta^2)s}{8\pi} \left(G_{\gamma\gamma}^V+2G_{Z\gamma}^V+G_{ZZ}^V+\frac{2\beta^2}{3-\beta^2}G_{ZZ}^A\right),\nonumber\\
A_{FB}^{\rm 0}&=\frac{-3\beta(H_{\gamma Z}+2H_{ZZ})}{(3-\beta^2)(G_{\gamma\gamma}^V+2G_{Z\gamma}^V+G_{ZZ}^V)+2\beta^2 G_{ZZ}^A},
\end{align}
with $\beta$ being the velocity of the top quark in the center-of-mass frame. The $G_{X,Y}^{V,A}$  and $H_{X,Y}$ factors are 
\bea
G_{X,Y}^{V,A} &=& \frac{g_e(X,Y)g_{V,A}^t(X) g_{V,A}^t(Y)}{\left(s-m_X^2\right) \left(s-m_Y^2\right)},\nonumber\\
H_{X,Y} &=& \frac{g_V^e(X) g_V^t(X) g_A^e(Y) g_A^t(Y)}{\left(s-m_X^2\right) \left(s-m_Y^2\right)},
\eea
where 
\bea
g_e(X,Y)=
\begin{cases}
g_V^e(\gamma)^2,  & X=Y=\gamma, \\
g_V^e(\gamma)g_V^e(Z), & X=\gamma, Y=Z, \\
g_V^e(Z)^2+g_A^e(Z)^2, & X=Y=Z.
\end{cases}
\eea
Here, $g_{V,A}^i(X/Y)$ denotes the vector ($V$) and axial-vector ($A$) coupling of gauge boson $X/Y=\gamma,Z$ to the electron ($i=e$) and top quark ($i=t$). In the SM,
\begin{align}
g_A^i(\gamma) &=0,  & g_V^i(\gamma) &=Q_i g s_W, \nonumber\\
g_A^i(Z) 	&= \frac{g}{c_W}g_A^i, & g_V^i(Z)&=\frac{g}{c_W}g_V^i,
\end{align}
where $Q_i$ is the electric charge of fermion $i$, $g_A^i=-T_3^i/2$ and $g_V^i=T_3^i/2-Q_i s_W^2$ with $T_3^e=-1/2$ and $T_3^t=1/2$, $s_W\equiv \sin\theta_W$ is the sine of the weak mixing angle.
The coefficients $a_V$, $a_A$, $b_V$ and $b_A$ are
\bea
a_V &=&\frac{2(G_{Z\gamma}^V+G_{ZZ}^V)}{g_V^t\left[G_{\gamma\gamma}^V+2G_{Z\gamma}^V+G_{ZZ}^V+2\beta^2/(3-\beta^2)G_{ZZ}^A\right]},\nonumber\\
a_A &=&\frac{4\beta^2 G_{ZZ}^A}
{g_A^t\left[(3-\beta^2)(G_{\gamma\gamma}^V+2G_{Z\gamma}^V+G_{ZZ}^V)+2\beta^2G_{ZZ}^A\right]},\nonumber\\
b_V &=&\frac{2H_{ZZ}}{g_V^t\left[H_{\gamma Z}+2H_{ZZ}\right]}-a_V,\nonumber\\
b_A &=&\frac{1}{g_A^t}-a_A.
\eea
The explicit expresses of $a_{L/R}$ and $b_{L/R}$ can be obtained from $a_{V/A}$ and $b_{V/A}$ as follows:
\bea 
a_L &=& \frac{1}{\left[G_{\gamma\gamma}^V+2G_{Z\gamma}^V+G_{ZZ}^V+2\beta^2/(3-\beta^2)G_{ZZ}^A\right]}
\Big(\dfrac{G_{Z\gamma}^V+G_{ZZ}^V}{g_V^t}-\dfrac{2\beta^2 G_{ZZ}^A}{g_A^t(3-\beta^2)}\Big),\nonumber \\
a_R &=& \frac{1}{2\left[G_{\gamma\gamma}^V+2G_{Z\gamma}^V+G_{ZZ}^V+2\beta^2/(3-\beta^2)G_{ZZ}^A\right]}
\Big(\dfrac{G_{Z\gamma}^V+G_{ZZ}^V}{g_V^t}+\dfrac{2\beta^2 G_{ZZ}^A}{g_A^t(3-\beta^2)}\Big),\nonumber \\
b_L &=&\frac{1}{2}(\frac{2H_{ZZ}}{g_V^t\left[H_{\gamma Z}+2H_{ZZ}\right]}-\frac{1}{g_A^t})-a_L,\nonumber \\
b_R &=&\frac{1}{4}(\frac{2H_{ZZ}}{g_V^t\left[H_{\gamma Z}+2H_{ZZ}\right]}+\frac{1}{g_A^t})-a_R.
\eea

\bibliographystyle{apsrev}
\bibliography{reference}

\end{document}